# Secure Beamforming in Full-Duplex SWIPT Systems

Yanjie Dong, Ahmed El Shafie, Md. Jahangir Hossain, Julian Cheng,
Naofal Al-Dhahir, and Victor C. M. Leung

**Abstract**

Physical layer security is a key issue in the full duplex (FD) communication systems due to the broadcast nature of wireless channels. In this paper, the joint design of information and artificial noise beamforming vectors is proposed for the FD simultaneous wireless information and power transferring (FD-SWIPT) systems. To guarantee high security and energy harvesting performance of the FD-SWIPT system, the proposed design is formulated as a sum information transmission rate (SITR) maximization problem under information-leakage and energy constraints. In addition, we consider the fairness issue between the uplink and downlink information transmission rates by formulating a fairness-aware SITR-maximization problem. Although the formulated SITR-maximization and fairness-aware SITR-maximization problems are non-convex, we solve them via semidefinite relaxation and one-dimensional search. The optimality of our proposed algorithms is theoretically proved, and the computation complexities are established. Moreover, we propose two suboptimal solutions to the formulated optimization problems. In terms of the SITR-maximization problem, numerical results show that the performance achieved by one of the two suboptimal algorithms is close to the performance of the optimal algorithm with increasing maximum transmission power of the FD-BST.

Yanjie Dong and Victor C. M. Leung are with the Department of Electrical and Computer Engineering, The University of British Columbia, Vancouver, BC V6T 1Z4, Canada (email:{ydong16, vleung}@ece.ubc.ca).

Ahmed El Shafie and Naofal Al-Dhahir are with the Electrical Engineering Department, The University of Texas at Dallas, Dallas, TX 75080, USA (e-mail: {ahmed.elshafie, aldhahir}@utdallas.edu).

Md. Jahangir Hossain and Julian Cheng are with the School of Engineering, The University of British Columbia, Kelowna, BC V1V 1V7, Canada (email:{jahangir.hossain, julian.cheng}@ubc.ca).

# I. Introduction

The upsurging wireless data volume drives the industrial and academic communities to search for efficient ways to boost the usage of the already scarce radio-frequency (RF) spectrum in the fifth generation (5G) wireless communication systems [1]. Full-duplex (FD) technology is considered as one of the promising solutions to improve the RF spectrum utilization due to its ability to double the spectrum efficiency. Moreover, energy efficiency is enhanced given the unchanged power consumption and improved spectrum efficency [1], [2]. However, FD nodes suffer from a strong loopback self-interference (LSI) due to the short distance between the transmitter and receiver sides on an FD node. Several LSI cancellation techniques are proposed in the FD literature, e.g., natural isolation [3], digital-analog domain cancellation [4], and spatial suppression [5]–[7]. By placing the absorptive shielding in the LSI channel, the natural isolation is attractive due to its implementation simplicity. However, the effectiveness of the natural isolation is limited by the form-factor of the wireless devices (i.e., the smaller the device, the less room to implement the isolator). It is even the case that using a high-end duplexer may not be sufficient for communication [5]. An alternative scheme to cancel the LSI is by using the digital and/or analog domain cancellation techniques. However, the dynamic range of the reception circuits is a bottleneck to cancel the strong LSI [2], [8]. As a result, spatial suppression, which is based on the spatial diversity to suppress the LSI, is required before further processing of the received signals [5]–[7].

Simultaneous wireless information and power transferring (SWIPT) systems are receiving increased attention [9]–[11]. Based on the energy carried by the RF signals, SWIPT systems can deliver information signal to the user equipments (UEs) and energy signal to the low energy-consumption energy receivers (ERs) [12]. Hence, the integration of FD technology into the SWIPT systems is promising due to the simultaneous support of UEs and ERs and the improvement of the spectrum efficiency and the energy efficiency [13]. To avoid the attenuation of the energy signal and meet energy harvesting demands, the ERs are usually deployed near the transmitters [14], [15]. However, the strong received signal strength at the ERs can also benefit eavesdropping when the ERs are malicious users. Therefore, there exists a critical tradeoff between the security issue and the energy requirements at the ERs in FD-SWIPT systems. It has been demonstrated in [16] that using part of the transmission power to artificially generate the noise, the information leakage to the eavesdroppers is degraded. Hence, the security in wireless communications is improved via broadcasting the artificial noise (AN) [16]–[19]. Motivated by this fact, we assume that information beamforming and AN vectors are jointly generated by the FD base station (FD-BST) and FD user equipment (FD-UE) to suppress the information leakage to the ERs. Another benefit of this setup is that the FD-BST and the FD-UE can coordinately perform AN beamforming, which improves the suppression efficiency of the information leakage [17], [20]–[22].

### *A. Related Work and Motivations*

Resource allocation is one of the key mechanisms to ensure the communication security in wireless communication systems, such as FD communication systems and the SWIPT systems. On the one hand, secure resource allocation in FD communication systems has been widely investigated for various topologies, e.g., point-to-point (PtP) topology [17], [18], [20], [23], point-to-multipoint (PtMP) topology [21], multipoint-to-multipoint (MPtMP) topology [24], and wireless relaying topology [7], [8], [17], [22]. Two major optimization problems are considered in secure FD communication systems, namely, secrecy-capacity maximization [17], [18], [22], [23] and transmission-power minimization [21]. For example, using jamming signals by the legitimate receiver, the authors in [17] studied secrecy-capacity maximization in FD communication systems with a PtP topology. With the same topology as [17], source and destination perform coordinated jamming in [18]. Using the secrecy-capacity region in [25], [26], the authors in [23] proposed an alternating optimization algorithm to obtain the locally optimal solution to maximize the secrecy capacity of the secure FD communication systems with a PtP topology.

On the other hand, secure resource allocation in the SWIPT systems was investigated in [13]–[15], [27]–[32]. To guarantee system security, different performance metrics are optimized, such as secrecy-capacity maximization [13], [14], [28], [31], security-outage-probability minimization [15], and transmission-power minimization [27], [32]. In SWIPT with a single legitimate UE, the authors in [14] and [28] investigated secrecy-capacity maximization for multiple single-antenna eavesdroppers and one single multiple-antenna eavesdropper, respectively. The authors in [32] investigated the power-minimization problem in SWIPT systems with multiple ERs and multiple multi-antenna eavesdroppers under the imperfect channel state information (CSI). In addition, the authors in [29] analyzed the secrecy capacity in SWIPT systems with a single legitimate UE and multiple eavesdroppers, and they obtained closed-form expressions for the security outage probability and the ergodic security capacity. The authors in [13] studied weighted-capacity maximization in the wireless relaying system, where the BST operates in the FD mode and a single ER can envisage energy and eavesdrop information from the received signals. Despite the plethora of research works on the security of wireless communication systems [13]–[15], [17]–[20], [23], [27]–[32], they are mainly based on the assumption that each eavesdropper can acquire information from one legitimate UE. Hence, the reported secrecy rates are, generally, not achievable when the ER can eavesdrop information from the simultaneous UE and BS transmissions [25], [26].

### B. Contributions

Motivated by the pioneering work on the capacity-region of FD two-way communication systems [25], [26], we consider a more general scenario, where there are multiple ERs in the FD-SWIPT system. In the FD-SWIPT system, the bidirectional communications between FD-BST and FD-UE place requirements on communication security and harvested energy at the multiple ERs. To the authors' best knowledge, the joint design of the information beamforming and AN vectors in the FD-SWIPT system with multiple ERs has not been reported in the current literature. Our objective is to maximize the sum rates for information transmissions of the FD-BST and FD-UE as well as confine the information leakage to the multiple ERs. Our main contributions are summarized as follows

- To avoid the saturation of the digital and analog circuits, we jointly design the information and AN vectors to suppress the LSI such that a certain information transmission rate is achieved.
- We investigate the maximization of the sum information transmission rates (SITR) in FD-SWIPT systems. Motivated by [14], we leverage the semidefinite relaxation (SDR) technique to develop an optimal solution to the SITR-maximization problem.
- We investigate the fairness-aware (FA) SITR-maximization problem, where the information transmission rates of the FD-BST and the FD-UE satisfy a predefined ratio. We perform convex relaxation to the proportional rate constraint and propose the FA-SITR algorithm to solve the relaxed optimization problem. We prove that the proposed FA-SITR algorithm converges to the optimal solution of the FA-SITR-maximization problem.

- In addition, we develop two suboptimal solutions with low complexity to each of the formulated problems: SITR-maximization and FA-SITR-maximization problems. For the SITR-maximization problem, we numerically show that the secrecy rate obtained by one of the proposed suboptimal algorithms converges to that of the optimal algorithm with an increasing maximum transmission power of the FD-BST.

Numerical results are used to verify the performance of the proposed algorithms. Specifically, we show that the optimal secrecy rate can be achieved by performing a two-dimensional search.

The remainder of this paper is organized as follows. In Section II, we describe the system model and problem formulation in FD-SWIPT systems. In Section III and Section IV, we respectively derive the optimal algorithms to the SITR-maximization and FA-SITR-maximization problems in FD-SWIPT systems. In Section V, we propose two suboptimal algorithms to each of the formulated problems and compare the computation complexities of the proposed optimal and suboptimal algorithms. Numerical results are presented in Section VI and the paper is concluded in Section VII.

*Notations:* Vectors and matrices are shown in bold lowercase letters and bold uppercase letters, respectively. $\mathbb{C}$ denotes the set of complex numbers. $\|\cdot\|_{\mathrm{F}}$ and $\|\cdot\|_p$ refer to the Frobenius norm and $\ell_p$-norm, respectively. $\sim$ stands for "distributed as". $\boldsymbol{I}$ denotes a identity matrix, and $\mathbf{0}_{N\times M}$ denotes a zero matrix with $N$ rows and $M$ columns. The expectation of a random variable is denoted as $\mathbb{E}\left[\cdot\right]$. The operator $(w)^+$ is defined as $(w)^+ \triangleq \max(w, 0)$. vec $[\cdot]$ converts an $M \times N$ matrix into a column vector of size $MN \times 1$. $\{\boldsymbol{w}_n\}_{n\in\mathcal{N}}$ represents the set composed of $\boldsymbol{w}_n$, $n \in \mathcal{N}$. For a square matrix $\boldsymbol{W}$, $\boldsymbol{W}^{\mathrm{H}}$ and $\mathrm{Tr}(\boldsymbol{W})$ denote its conjugate transpose and trace, respectively. $\boldsymbol{W} \succeq \mathbf{0}$ and $\boldsymbol{W} \succ \mathbf{0}$ imply that $\boldsymbol{W}$ is a positive semidefinite and a positive definite matrix, respectively.

## II. System Model and Problem Formulation

### A. System Model and Assumptions

We consider an FD-SWIPT system which consists of an FD-BST, a legitimate FD-UE, and $K$ ERs. Let $\mathcal{K} = \{1, 2, \ldots, K\}$ be the set of ERs, where each ER can also be the potential eavesdroppers. The FD-BST is equipped with $N_a > 1$ transmit antennas and a single receive antenna for simultaneous transmission and reception over the same frequency band. Similarly, the legitimate FD-UE is equipped with $N_b > 1$ transmit antennas and a single receive antenna.

Thus, the FD-BST and the FD-UE operate in FD mode. The $k$-th ER is equipped with a single receive antenna. We consider frame-based communication over frequency-nonselective fading channels with unit duration for each frame. Therefore, the terms "power" and "energy" can be used interchangeably.

We assume that the FD-BST and the FD-UE can perfectly estimate the CSI [7]. This assumption is reasonable since the ERs and the FD-UE are assumed to be non-hostile legitimate nodes in the FD-SWIPT system. However, the ERs are curious to decode the information signals transmitted by the FD-BST and the FD-UE. Therefore, the ERs are assumed to be the potential eavesdroppers. At the beginning of each frame, each node sends a pilot signal to the FD-BST. All nodes listen to the pilot signals and estimate the channels associated with the node sending the pilot signal. After that, the FD-BST sends another pilot signal such that each node estimates the channel connected to the FD-BST. Then, the FD-UE feeds back the estimated CSI to the FD-BST. Finally, the FD-BST and the FD-UE use the beamforming vectors for information transmission and the AN signals for coordinated jamming.

The transmission signals of the FD-BST and FD-UE are, respectively, given by

$$\boldsymbol{x}_{a,b} = \boldsymbol{w}_{a,b} s_b + \boldsymbol{v}_a \tag{1}$$

and

$$\boldsymbol{x}_{b,a} = \boldsymbol{w}_{b,a} s_a + \boldsymbol{v}_b \tag{2}$$

where $s_a \sim \mathcal{CN}(0,1)$ and $s_b \sim \mathcal{CN}(0,1)$ denote, respectively, the information-bearing signals for the FD-BST and the FD-UE, and $s_a$ and $s_b$ are independent from each other. The vectors $\boldsymbol{w}_{a,b} \in \mathbb{C}^{N_a \times 1}$ and $\boldsymbol{v}_a \in \mathbb{C}^{N_a \times 1}$ ($\boldsymbol{w}_{b,a} \in \mathbb{C}^{N_b \times 1}$ and $\boldsymbol{v}_b \in \mathbb{C}^{N_b \times 1}$) are, respectively, the information beamforming vector and the AN vector of the FD-BST (FD-UE). In particular, the AN vector $\boldsymbol{v}_a$ ($\boldsymbol{v}_b$) is modeled as a circularly symmetric complex Gaussian vector with mean zero and covariance $\boldsymbol{V}_{a,a} \succeq 0$ ($\boldsymbol{V}_{b,b} \succeq 0$).

Therefore, the received signal at the $k$-th ER is given by

$$y_{e_k} = \boldsymbol{h}_{a,e_k}^{\mathrm{H}} \boldsymbol{w}_{a,b} s_b + \boldsymbol{h}_{b,e_k}^{\mathrm{H}} \boldsymbol{w}_{b,a} s_a + \boldsymbol{h}_{e_k}^{\mathrm{H}} \boldsymbol{v} + z_{e_k} \tag{3}$$

where $\boldsymbol{h}_{e_k} \triangleq \mathrm{vec}\left([\boldsymbol{h}_{a,e_k}, \boldsymbol{h}_{b,e_k}]\right)$ with the channel coefficient vectors from the FD-BST and the FD-UE to the $k$-th ER denoted by $\boldsymbol{h}_{a,e_k} \in \mathbb{C}^{N_a \times 1}$ and $\boldsymbol{h}_{b,e_k} \in \mathbb{C}^{N_b \times 1}$, respectively. The term $\boldsymbol{z}_{e_k} \sim \mathcal{CN}\left(0, \sigma_{e_k}^2\right)$ denotes the additive white Gaussian noise (AWGN) at the $k$-th ER with zero mean and variance $\sigma_{e_k}^2$; $\boldsymbol{v} \triangleq \mathrm{vec}\left([\boldsymbol{v}_a, \boldsymbol{v}_b]\right)$ is the compact form of the coordinated AN vector[1]. Here, the AN vector $\boldsymbol{v}$ is modeled as a circularly symmetric Gaussian vector with zero mean and covariance matrix $\boldsymbol{V}$. The AN covariance matrix $\boldsymbol{V}$ is given by

$$\boldsymbol{V} = \begin{bmatrix} \boldsymbol{V}_{a,a} & \boldsymbol{V}_{a,b} \\ \boldsymbol{V}_{b,a} & \boldsymbol{V}_{b,b} \end{bmatrix} \tag{4}$$

where the matrices $\boldsymbol{V}_{a,b}$ and $\boldsymbol{V}_{b,a}$ are defined as $\boldsymbol{V}_{a,b} = \mathbb{E}\left[\boldsymbol{v}_a \boldsymbol{v}_b^{\mathrm{H}}\right]$ and $\boldsymbol{V}_{b,a} = \mathbb{E}\left[\boldsymbol{v}_b \boldsymbol{v}_a^{\mathrm{H}}\right]$, respectively.

The achievable sum data rate of the $k$-th ER is given by

$$C_{e_k}\left(\boldsymbol{W}_{a,b}, \boldsymbol{W}_{b,a}, \boldsymbol{V}\right) = \log\left(1 + \frac{\mathrm{Tr}\left(\boldsymbol{H}_{a,e_k}\boldsymbol{W}_{a,b}\right) + \mathrm{Tr}\left(\boldsymbol{H}_{b,e_k}\boldsymbol{W}_{b,a}\right)}{\mathrm{Tr}\left(\boldsymbol{H}_{e_k}\boldsymbol{V}\right) + \sigma_{e_k}^2}\right). \tag{5}$$

In addition, the following two achievable data rates of the $k$-th ER are given by

$$C_{a,e_k}\left(\boldsymbol{W}_{a,b}, \boldsymbol{W}_{b,a}, \boldsymbol{V}\right) = \log\left(1 + \frac{\mathrm{Tr}\left(\boldsymbol{H}_{b,e_k}\boldsymbol{W}_{b,a}\right)}{\mathrm{Tr}\left(\boldsymbol{H}_{a,e_k}\boldsymbol{W}_{a,b}\right) + \mathrm{Tr}\left(\boldsymbol{H}_{e_k}\boldsymbol{V}\right) + \sigma_{e_k}^2}\right), \forall k \tag{6}$$

and

$$C_{b,e_k}\left(\boldsymbol{W}_{a,b}, \boldsymbol{W}_{b,a}, \boldsymbol{V}\right) = \log\left(1 + \frac{\mathrm{Tr}\left(\boldsymbol{H}_{a,e_k}\boldsymbol{W}_{a,b}\right)}{\mathrm{Tr}\left(\boldsymbol{H}_{b,e_k}\boldsymbol{W}_{b,a}\right) + \mathrm{Tr}\left(\boldsymbol{H}_{e_k}\boldsymbol{V}\right) + \sigma_{e_k}^2}\right), \forall k \tag{7}$$

where $\boldsymbol{W}_{a,b} \triangleq \boldsymbol{w}_{a,b}\boldsymbol{w}_{a,b}^{\mathrm{H}}$, $\boldsymbol{W}_{b,a} \triangleq \boldsymbol{w}_{b,a}\boldsymbol{w}_{b,a}^{\mathrm{H}}$, $\boldsymbol{H}_{a,e_k} \triangleq \boldsymbol{h}_{a,e_k}\boldsymbol{h}_{a,e_k}^{\mathrm{H}}$, $\boldsymbol{H}_{b,e_k} \triangleq \boldsymbol{h}_{b,e_k}\boldsymbol{h}_{b,e_k}^{\mathrm{H}}$ and $\boldsymbol{H}_{e_k} \triangleq \boldsymbol{h}_{e_k}\boldsymbol{h}_{e_k}^{\mathrm{H}}$. The ranks of beamforming matrices $\boldsymbol{W}_{a,b}$ and $\boldsymbol{W}_{b,a}$ are upper-bounded as

$$\mathrm{Rank}\left(\boldsymbol{W}_{a,b}\right) \leq 1 \text{ and } \mathrm{Rank}\left(\boldsymbol{W}_{b,a}\right) \leq 1. \tag{8}$$

The amount of harvested energy at the $k$-th ER is given by

$$E_k = \eta\left(\mathrm{Tr}\left(\boldsymbol{H}_{e_k}\boldsymbol{V}\right) + \mathrm{Tr}\left(\boldsymbol{H}_{a,e_k}\boldsymbol{W}_{a,b}\right) + \mathrm{Tr}\left(\boldsymbol{H}_{b,e_k}\boldsymbol{W}_{b,a}\right) + \sigma_{e_k}^2\right) \tag{9}$$

where $0 \leq \eta \leq 1$ is the efficiency of the energy harvester.

In addition, the received signals at the FD-BST and the FD-UE are, respectively, given by

$$y_a = \boldsymbol{h}_{b,a}^{\mathrm{H}}\boldsymbol{w}_{b,a}s_a + \boldsymbol{h}_{a,a}^{\mathrm{H}}\boldsymbol{w}_{a,b}s_b + \boldsymbol{h}_a^{\mathrm{H}}\boldsymbol{v} + z_a \tag{10}$$

[1]Using the channel matrices between the legitimate nodes, the AN symbols are generated from a pseudo-random signal which is perfectly known at the legitimate nodes but not at the eavesdroppers. This is realized by using a short secret key as a seed information for the Gaussian pseudo-random sequence generator. The legitimate nodes regularly change the secret key seeds to maintain the AN sequence secured from the eavesdropper. A similar assumption can be found in, e.g., [20], [33] and the references therein.

and

$$y_b = \boldsymbol{h}_{a,b}^{\mathrm{H}} \boldsymbol{w}_{a,b} s_b + \boldsymbol{h}_{b,b}^{\mathrm{H}} \boldsymbol{w}_{b,a} s_a + \boldsymbol{h}_b^{\mathrm{H}} \boldsymbol{v} + z_b \tag{11}$$

where $z_a \sim \mathcal{CN}\left(0, \sigma_a^2\right)$ and $z_b \sim \mathcal{CN}\left(0, \sigma_b^2\right)$ denote, respectively, the AWGN at the FD-BST and FD-UE; $\boldsymbol{h}_a \triangleq \mathrm{vec}\left(\left[\boldsymbol{h}_{a,a}, \boldsymbol{h}_{b,a}\right]\right)$ and $\boldsymbol{h}_b \triangleq \mathrm{vec}\left(\left[\boldsymbol{h}_{a,b}, \boldsymbol{h}_{b,b}\right]\right)$ denote the compact form of the channel coefficient vectors for the FD-BST and the FD-UE, respectively. Without loss of generality, we assume that the channel coefficient vectors $\boldsymbol{h}_a$, $\boldsymbol{h}_b$ and $\boldsymbol{h}_{e_k}$ are statistically independent.

With the rank constraints in (8), the information transmission rate of the FD-UE and the FD-BST are, respectively, given by

$$C_a\left(\boldsymbol{W}_{a,b}, \boldsymbol{W}_{b,a}, \boldsymbol{V}\right) = \log\left(1 + \frac{\mathrm{Tr}\left(\boldsymbol{H}_{b,a}\boldsymbol{W}_{b,a}\right)}{\mathrm{Tr}\left(\boldsymbol{H}_{a,a}\boldsymbol{W}_{a,b}\right) + \mathrm{Tr}\left(\boldsymbol{H}_a\boldsymbol{V}\right) + \sigma_a^2}\right) \tag{12}$$

and

$$C_b\left(\boldsymbol{W}_{a,b}, \boldsymbol{W}_{b,a}, \boldsymbol{V}\right) = \log\left(1 + \frac{\mathrm{Tr}\left(\boldsymbol{H}_{a,b}\boldsymbol{W}_{a,b}\right)}{\mathrm{Tr}\left(\boldsymbol{H}_{b,b}\boldsymbol{W}_{b,a}\right) + \mathrm{Tr}\left(\boldsymbol{H}_b\boldsymbol{V}\right) + \sigma_b^2}\right) \tag{13}$$

where $\boldsymbol{H}_a \triangleq \boldsymbol{h}_a\boldsymbol{h}_a^{\mathrm{H}}$ $\boldsymbol{H}_{a,a} \triangleq \boldsymbol{h}_{a,a}\boldsymbol{h}_{a,a}^{\mathrm{H}}$, $\boldsymbol{H}_{a,b} \triangleq \boldsymbol{h}_{a,b}\boldsymbol{h}_{a,b}^{\mathrm{H}}$, $\boldsymbol{H}_b \triangleq \boldsymbol{h}_b\boldsymbol{h}_b^{\mathrm{H}}$, $\boldsymbol{H}_{b,b} \triangleq \boldsymbol{h}_{b,b}\boldsymbol{h}_{b,b}^{\mathrm{H}}$, and $\boldsymbol{H}_{b,a} \triangleq \boldsymbol{h}_{b,a}\boldsymbol{h}_{b,a}^{\mathrm{H}}$. Note that the signals received by the FD-BST and FD-UE contain the confidential information and the auxiliary information [26].

Based on Lemma 1 in [26], the confidential information leaked to the ERs is zero if the following inequalities are satisfied

$$C_{a,e_k}\left(\boldsymbol{W}_{a,b}, \boldsymbol{W}_{b,a}, \boldsymbol{V}\right) \leq C_a^{\mathrm{LEAK}}, \forall k \tag{14}$$

$$C_{b,e_k}\left(\boldsymbol{W}_{a,b}, \boldsymbol{W}_{b,a}, \boldsymbol{V}\right) \leq C_b^{\mathrm{LEAK}}, \forall k \tag{15}$$

$$C_{e_k}\left(\boldsymbol{W}_{a,b}, \boldsymbol{W}_{b,a}, \boldsymbol{V}\right) \leq C^{\mathrm{LEAK}}, \forall k \tag{16}$$

where $C_a^{\mathrm{LEAK}}$ and $C_b^{\mathrm{LEAK}}$ are, respectively, the information-leakage rate thresholds of the FD-BST and FD-UE; $C^{\mathrm{LEAK}}$ denotes the threshold for the sum information-leakage rate of the FD-BST and FD-UE. Based on the Corollary 1 in [26], the secrecy rate of the FD-SWIPT system is obtained as

$$C_a\left(\boldsymbol{W}_{a,b}, \boldsymbol{W}_{b,a}, \boldsymbol{V}\right) + C_b\left(\boldsymbol{W}_{a,b}, \boldsymbol{W}_{b,a}, \boldsymbol{V}\right) - \max\left\{C_a^{\mathrm{LEAK}} + C_b^{\mathrm{LEAK}}, C^{\mathrm{LEAK}}\right\}. \tag{17}$$

***Remark 1:*** From (17), we observe that the secrecy rate is related to the information transmission rates $C_a\left(\boldsymbol{W}_{a,b}, \boldsymbol{W}_{b,a}, \boldsymbol{V}\right)$ and $C_b\left(\boldsymbol{W}_{a,b}, \boldsymbol{W}_{b,a}, \boldsymbol{V}\right)$ as well as the information leakage rates $C_{e_k}\left(\boldsymbol{W}_{a,b}, \boldsymbol{W}_{b,a}, \boldsymbol{V}\right)$, $C_{a,e_k}\left(\boldsymbol{W}_{a,b}, \boldsymbol{W}_{b,a}, \boldsymbol{V}\right)$ and $C_{b,e_k}\left(\boldsymbol{W}_{a,b}, \boldsymbol{W}_{b,a}, \boldsymbol{V}\right)$. Moreover, the achievable rates at all nodes and the harvested energy at the ERs are controlled by the selection of the information and AN beamforming vectors. With the appropriate design of the beamforming and the AN vectors, the secrecy rate can be maximized when the summation of the information transmission rate is maximized with confined information leakage to the ERs. In this work, we investigate the sum information transmission rate (SITR) maximization subjected to the information leakage constraints and energy harvesting constraints.

### B. Problem Formulation

Our objective is to maximize the summation of information transmission rate of the FD-SWIPT system subjected to the information-leakage constraints, the maximum transmission power constraints and the required harvested energy of each ER. Therefore, the SITR-maximization problem can be formulated as

$$\max_{\boldsymbol{W}_{a,b}, \boldsymbol{W}_{b,a}, \boldsymbol{V}} C_a\left(\boldsymbol{W}_{a,b}, \boldsymbol{W}_{b,a}, \boldsymbol{V}\right) + C_b\left(\boldsymbol{W}_{a,b}, \boldsymbol{W}_{b,a}, \boldsymbol{V}\right) \tag{18a}$$

$$\text{s.t. } C_{a,e_k}\left(\boldsymbol{W}_{a,b}, \boldsymbol{W}_{b,a}, \boldsymbol{V}\right) \leq C_a^{\text{LEAK}}, \forall k \tag{18b}$$

$$C_{b,e_k}\left(\boldsymbol{W}_{a,b}, \boldsymbol{W}_{b,a}, \boldsymbol{V}\right) \leq C_b^{\text{LEAK}}, \forall k \tag{18c}$$

$$C_{e_k}\left(\boldsymbol{W}_{a,b}, \boldsymbol{W}_{b,a}, \boldsymbol{V}\right) \leq C^{\text{LEAK}}, \forall k \tag{18d}$$

$$\text{Tr}\left(\boldsymbol{B}_a \boldsymbol{V}\right) + \text{Tr}\left(\boldsymbol{W}_{a,b}\right) \leq P_a^{\max} \tag{18e}$$

$$\text{Tr}\left(\boldsymbol{B}_b \boldsymbol{V}\right) + \text{Tr}\left(\boldsymbol{W}_{b,a}\right) \leq P_b^{\max} \tag{18f}$$

$$\text{Tr}\left(\boldsymbol{H}_{e_k} \boldsymbol{V}\right) + \text{Tr}\left(\boldsymbol{H}_{a,e_k} \boldsymbol{W}_{a,b}\right) + \text{Tr}\left(\boldsymbol{H}_{b,e_k} \boldsymbol{W}_{b,a}\right) + \sigma_{e_k}^2 \geq \frac{P_k^{\text{REQ}}}{\eta}, \forall k \tag{18g}$$

$$\boldsymbol{W}_{a,b} \succeq 0, \boldsymbol{W}_{b,a} \succeq 0, \boldsymbol{V} \succeq 0 \tag{18h}$$

$$\text{Rank}\left(\boldsymbol{W}_{a,b}\right) \leq 1 \text{ and } \text{Rank}\left(\boldsymbol{W}_{b,a}\right) \leq 1 \tag{18i}$$

where $P_a^{\max}$ and $P_b^{\max}$ are maximum transmission power for the FD-BST and the FD-UE, respectively. $P_k^{\text{req}}$ denotes the required amount of harvested energy by the $k$-th ER. Here, the matrices $\boldsymbol{B}_a$ and $\boldsymbol{B}_b$ are, respectively, defined as

$$\boldsymbol{B}_a = \begin{bmatrix} \boldsymbol{I} & \mathbf{0}_{N_a \times N_b} \\ \mathbf{0}_{N_b \times N_a} & \mathbf{0}_{N_b \times N_b} \end{bmatrix} \tag{19}$$

and

$$\boldsymbol{B}_b = \begin{bmatrix} \mathbf{0}_{N_a \times N_a} & \mathbf{0}_{N_a \times N_b} \\ \mathbf{0}_{N_b \times N_a} & \boldsymbol{I} \end{bmatrix}. \tag{20}$$

Since the fairness issue between the FD-BST and FD-UE is important, we also investigate the SITR-maximization problem with a fairness constraint. The FA-SITR-maximization problem is formulated as

$$\max_{\boldsymbol{W}_{a,b}, \boldsymbol{W}_{b,a}, \boldsymbol{V}} C_a\left(\boldsymbol{W}_{a,b}, \boldsymbol{W}_{b,a}, \boldsymbol{V}\right) + C_b\left(\boldsymbol{W}_{a,b}, \boldsymbol{W}_{b,a}, \boldsymbol{V}\right) \tag{21a}$$

$$\text{s.t. (18b)} - \text{(18i)} \tag{21b}$$

$$\frac{C_a\left(\boldsymbol{W}_{a,b}, \boldsymbol{W}_{b,a}, \boldsymbol{V}\right)}{C_b\left(\boldsymbol{W}_{a,b}, \boldsymbol{W}_{b,a}, \boldsymbol{V}\right)} = \frac{\phi_a}{\phi_b} \tag{21c}$$

where $\phi_a \geq 0$ and $\phi_b \geq 0$ are the fractions of the information transmission rate of the FD-UE and the FD-BST with $\phi_a + \phi_b = 1$. As the ratio $\phi_a/\phi_b$ increases (decreases), more emphasize is on the information transmission rate of the FD-UE (FD-BST). By controlling this ratio, we can achieve any required fairness between the two information rates.

## III. SITR Maximization

### A. Feasibility Analysis

The SITR-maximization problem (18) can be infeasible under certain channel conditions when the values of $P_a^{\max}$ and $P_b^{\max}$ are too low and/or the value of $P_k^{\text{req}}$ is too high, $k \in \mathcal{K}$. By setting the objective value of (18) equal to zero, the FD-BST and the FD-UE do not need to transmit information. Therefore, the beamforming matrices $\boldsymbol{W}_{a,b}$ and $\boldsymbol{W}_{b,a}$ are the zero matrices. The FD-BST and the FD-UE only need to transmit the AN vectors to satisfy the energy requirements. When the transmission power of the FD-BST and the FD-UE cannot satisfy the energy requirements of the ERs, the SITR-maximization problem (18) is infeasible. Hence, the feasibility of the optimization problem (18) can be checked by solving the following problem

$$\begin{aligned} &\text{Find } \boldsymbol{V} \succeq \mathbf{0} \\ &\text{s.t. } \operatorname{Tr}\left(\boldsymbol{B}_a \boldsymbol{V}\right) \leq P_a^{\max} \\ &\quad \operatorname{Tr}\left(\boldsymbol{B}_b \boldsymbol{V}\right) \leq P_b^{\max} \\ &\quad \operatorname{Tr}\left(\boldsymbol{H}_{e_k} \boldsymbol{V}\right) + \sigma_{e_k}^2 \geq \frac{P_k^{\text{REQ}}}{\eta}, \forall k. \end{aligned} \tag{22}$$

The feasible region of (22) is convex; therefore, we can perform the feasibility check via CVX [34]. If the matrix $\boldsymbol{V}$ is obtained via solving (22), we have $(\boldsymbol{W}_{a,b} = \boldsymbol{0}, \boldsymbol{W}_{b,a} = \boldsymbol{0}, \boldsymbol{V})$ is a feasible solution to the SITR-maximization problem (18). Without loss of generality, we assume the SITR-maximization problem (18) is feasible hereinafter.

### *B. Optimal Joint Beamforming and AN Vectors to SITR-Maximization Problem*

Since the SITR-maximization problem (18) is non-convex due to the rank-one constraints in (18i) and the non-convex objective function, we develop develop a two-stage algorithm based on the semidefinite relaxation (SDR) technique and one-dimensional search in this subsection. Introducing a parameter $\beta$, we recast the optimization problem (18) as

$$\max_{\boldsymbol{W}_{a,b},\boldsymbol{W}_{b,a},\boldsymbol{V}} \frac{\mathrm{Tr}\left(\boldsymbol{H}_{b,a}\boldsymbol{W}_{b,a}\right)}{\mathrm{Tr}\left(\boldsymbol{H}_{a,a}\boldsymbol{W}_{a,b}\right) + \mathrm{Tr}\left(\boldsymbol{H}_{a}\boldsymbol{V}\right) + \sigma_a^2} \tag{23a}$$

$$\text{s.t.} \ \frac{\mathrm{Tr}\left(\boldsymbol{H}_{b,e_k}\boldsymbol{W}_{b,a}\right)}{\exp\left(C_a^{\text{LEAK}}\right) - 1} \le \mathrm{Tr}\left(\boldsymbol{H}_{a,e_k}\boldsymbol{W}_{a,b}\right) + \mathrm{Tr}\left(\boldsymbol{H}_{e_k}\boldsymbol{V}\right) + \sigma_{e_k}^2, \forall k \tag{23b}$$

$$\frac{\mathrm{Tr}\left(\boldsymbol{H}_{a,e_k}\boldsymbol{W}_{a,b}\right)}{\exp\left(C_b^{\text{LEAK}}\right) - 1} \le \mathrm{Tr}\left(\boldsymbol{H}_{b,e_k}\boldsymbol{W}_{b,a}\right) + \mathrm{Tr}\left(\boldsymbol{H}_{e_k}\boldsymbol{V}\right) + \sigma_{e_k}^2, \forall k \tag{23c}$$

$$\frac{\mathrm{Tr}\left(\boldsymbol{H}_{a,e_k}\boldsymbol{W}_{a,b}\right) + \mathrm{Tr}\left(\boldsymbol{H}_{b,e_k}\boldsymbol{W}_{b,a}\right)}{\exp\left(C^{\text{LEAK}}\right) - 1} \le \mathrm{Tr}(\boldsymbol{H}_{e_k}\boldsymbol{V}) + \sigma_{e_k}^2, \forall k \tag{23d}$$

$$\text{(18e)} - \text{(18i)} \tag{23e}$$

where $\beta \in \left[0, \log\left(1 + P_a^{\max}\sqrt{\mathrm{Tr}\left(\boldsymbol{H}_{a,b}\right)}/\sigma_b^2\right)\right]$.

Thus, the objective value of the SITR-maximization problem (18) is obtained as

$$\begin{aligned} &\max_{\beta} \ \log\left(1 + \mathcal{OBJ}_{\text{SITR}}\left(\beta\right)\right) + \beta \\ &\text{s.t.} \ \beta \in \left[0, \log\left(1 + P_a^{\max}\sqrt{\mathrm{Tr}\left(\boldsymbol{H}_{a,b}\right)}/\sigma_b^2\right)\right] \end{aligned} \tag{24}$$

where $\mathcal{OBJ}_{\text{SITR}}\left(\beta\right)$ is the objective value of (23). Hence, the procedures to obtain the optimal value of the optimization problem (18) are summarized as: 1) solving the optimization problem (23) given a fixed value for $\beta$; then 2) performing golden-section search for the optimal value of $\beta \in \left[0, \log\left(1 + P_a^{\max}\sqrt{\mathrm{Tr}\left(\boldsymbol{H}_{a,b}\right)}/\sigma_b^2\right)\right]$. Moreover, the system secrecy rate is obtained as $\log\left(1 + \mathcal{OBJ}_{\text{SITR}}\left(\beta\right)\right) + \beta - \max\left\{C_a^{\text{LEAK}} + C_b^{\text{LEAK}}, C^{\text{LEAK}}\right\}$.

We observe that the optimization problem (23) is still non-convex due to the linear fractional objective function and the rank-one constraints. An intuitive method to deal with the fractional objective function is by leveraging the Dinkelbach transformation and iteratively solving a series of convex-optimization problems [35], [36]. However, the Dinkelbach transformation generally requires several iterations and a large amount of computation resources. Hence, we are motivated to use the Charnes-Cooper transformation such that the optimal solution is obtained without the iterations introduced by the Dinkelbach transformation [14], [37].

Introducing an auxiliary variable $\gamma \geq 0$ and dropping the rank-one constraints in (18i), we obtain the SDR version of the optimization problem (23) as

$$\min_{\boldsymbol{W}_{a,b},\boldsymbol{W}_{b,a},\boldsymbol{V},\gamma} \ -\mathrm{Tr}\left(\boldsymbol{H}_{b,a}\boldsymbol{W}_{b,a}\right) \tag{25a}$$

$$\text{s.t. } \mathrm{Tr}\left(\boldsymbol{H}_{a,a}\boldsymbol{W}_{a,b}\right)+\mathrm{Tr}\left(\boldsymbol{H}_{a}\boldsymbol{V}\right)+\gamma\sigma_a^2=1 \tag{25b}$$

$$\frac{\mathrm{Tr}\left(\boldsymbol{H}_{b,e_k}\boldsymbol{W}_{b,a}\right)}{\exp\left(C_a^{\text{LEAK}}\right)-1} \leq \mathrm{Tr}\left(\boldsymbol{H}_{a,e_k}\boldsymbol{W}_{a,b}\right)+\mathrm{Tr}\left(\boldsymbol{H}_{e_k}\boldsymbol{V}\right)+\gamma\sigma_{e_k}^2, \forall k \tag{25c}$$

$$\frac{\mathrm{Tr}\left(\boldsymbol{H}_{a,e_k}\boldsymbol{W}_{a,b}\right)}{\exp\left(C_b^{\text{LEAK}}\right)-1} \leq \mathrm{Tr}\left(\boldsymbol{H}_{b,e_k}\boldsymbol{W}_{b,a}\right)+\mathrm{Tr}\left(\boldsymbol{H}_{e_k}\boldsymbol{V}\right)+\gamma\sigma_{e_k}^2, \forall k \tag{25d}$$

$$\frac{\mathrm{Tr}\left(\boldsymbol{H}_{a,e_k}\boldsymbol{W}_{a,b}\right)+\mathrm{Tr}\left(\boldsymbol{H}_{b,e_k}\boldsymbol{W}_{b,a}\right)}{\exp\left(C^{\text{LEAK}}\right)-1} \leq \mathrm{Tr}(\boldsymbol{H}_{e_k}\boldsymbol{V})+\gamma\sigma_{e_k}^2, \forall k \tag{25e}$$

$$\frac{\mathrm{Tr}\left(\boldsymbol{H}_{a,b}\boldsymbol{W}_{a,b}\right)}{\exp\left(\beta\right)-1} \geq \mathrm{Tr}\left(\boldsymbol{H}_{b,b}\boldsymbol{W}_{b,a}\right)+\mathrm{Tr}\left(\boldsymbol{H}_{b}\boldsymbol{V}\right)+\gamma\sigma_b^2 \tag{25f}$$

$$\mathrm{Tr}\left(\boldsymbol{B}_{a}\boldsymbol{V}\right)+\mathrm{Tr}\left(\boldsymbol{W}_{a,b}\right) \leq \gamma P_a^{\max} \tag{25g}$$

$$\mathrm{Tr}\left(\boldsymbol{B}_{b}\boldsymbol{V}\right)+\mathrm{Tr}\left(\boldsymbol{W}_{b,a}\right) \leq \gamma P_b^{\max} \tag{25h}$$

$$\mathrm{Tr}\left(\boldsymbol{H}_{a,e_k}\boldsymbol{W}_{a,b}\right)+\mathrm{Tr}\left(\boldsymbol{H}_{b,e_k}\boldsymbol{W}_{b,a}\right)+\mathrm{Tr}\left(\boldsymbol{H}_{e_k}\boldsymbol{V}\right) \geq \gamma\left(\frac{P_k^{\text{REQ}}}{\eta}-\sigma_{e_k}^2\right), \forall k \tag{25i}$$

$$\boldsymbol{W}_{a,b} \succeq \boldsymbol{0}, \boldsymbol{W}_{b,a} \succeq \boldsymbol{0}, \boldsymbol{V} \succeq \boldsymbol{0}, \gamma \geq 0. \tag{25j}$$

If an optimal solution to (25) satisfies the rank-one constraints, the optimal solution to (25) is also an optimal solution to (23), which is verified in Appendix A. Since the inequality constraints in the optimization problem (25) are affine, the Slater's condition holds when the optimization problem (25) is feasible [38]. In addition, the optimization problem has an affine objective function; therefore, the duality gap of the optimization problem (25) is zero.

***Proposition** 1:* Assuming that the optimization problem (25) is feasible, there exists an optimal solution $\left(\boldsymbol{W}_{a,b}^*, \boldsymbol{W}_{b,a}^*, \boldsymbol{V}^*, \gamma^*\right)$ to (25) that satisfies the rank-one constraints for $\boldsymbol{W}_{a,b}^*$ and $\boldsymbol{W}_{b,a}^*$, i.e., Rank $\left(\boldsymbol{W}_{a,b}^*\right) \leq 1$ and Rank $\left(\boldsymbol{W}_{b,a}^*\right) \leq 1$.

*Proof:* See Appendix B. □

The proof of Proposition 1 shows that the optimal objective value of (23) can be obtained via solving the optimization problem (25) and performing the rank-one recovery procedures in (60)-(62). Based on the golden-section search for $\beta$, we summarize the procedure for solving the SITR-maximization problem (18) in Algorithm 1.

**Algorithm 1** SITR Algorithm

1: FD-BST sets the iteration index $\tau = 0$ and the stop threshold $\epsilon$
2: FD-BST obtains $\beta^{\min} = 0$ and $\beta^{\max} = \log\left(1 + P_a^{\max}\sqrt{\mathrm{Tr}\left(\boldsymbol{H}_{a,b}\right)}/\sigma_b^2\right)$.
3: **repeat**
4: FD-BST obtains $\beta_1 = \beta^{\min} + 0.382\left(\beta^{\max} - \beta^{\min}\right)$ and $\beta_2 = \beta^{\max} - 0.382\left(\beta^{\max} - \beta^{\min}\right)$
5: FD-BST obtains the objective values $\mathcal{OBJ}_{\text{SITR}}\left(\beta_1\right)$ and $\mathcal{OBJ}_{\text{SITR}}\left(\beta_2\right)$ via solving (25)
6: **if** $\mathcal{OBJ}_{\text{SITR}}\left(\beta_1\right) \geq \mathcal{OBJ}_{\text{SITR}}\left(\beta_2\left(\tau\right)\right)$ **then**
7: $\beta^{\max} \leftarrow \beta_2$
8: **else**
9: $\beta^{\min} \leftarrow \beta_1$
10: **end if**
11: **until** $\left|\beta^{\max} - \beta^{\min}\right| \leq \epsilon$
12: FD-BST obtains the minimum value of (18) and the beamforming matrices $\boldsymbol{W}_{a,b}^{*}$, $\boldsymbol{W}_{b,a}^{*}$, $\boldsymbol{V}^{*}$
13: **if** Rank $\left(\boldsymbol{W}_{a,b}^{*}\right) > 1$ and/or Rank $\left(\boldsymbol{W}_{b,a}^{*}\right) > 1$ **then**
14: FD-BST recovers the rank-one constraints via (60)-(62) and obtains $\widetilde{\boldsymbol{W}}_{a,b}^{*}$, $\widetilde{\boldsymbol{W}}_{b,a}^{*}$ and $\widetilde{\boldsymbol{V}}^{*}$.
15: **end if**
16: FD-BST performs the eigenvalue decomposition for matrices $\widetilde{\boldsymbol{W}}_{a,b}^{*}$, $\widetilde{\boldsymbol{W}}_{b,a}^{*}$ and $\widetilde{\boldsymbol{V}}^{*}$ in order to obtain the information beamforming vectors $\boldsymbol{w}_{a,b}^{*}$ and $\boldsymbol{w}_{b,a}^{*}$ and the AN vectors $\boldsymbol{v}_a^{*}$ and $\boldsymbol{v}_b^{*}$.
17: FD-BST notifies the FD-UE the information beamforming vector $\boldsymbol{w}_{b,a}^{*}$ and AN vector $\boldsymbol{v}_b^{*}$

*Complexity Analysis of the SITR Algorithm:* We observe that the major complexity of the SITR algorithm lies in the iteration loop between lines 3–11. Hence, we use the complexity of the iteration loop to evaluate the complexity of the SITR algorithm. In each iteration, the optimization problem (25) is solved twice. Moreover, the optimization problem (25) contains three matrix variables, one nonnonnegative variable, $4+4K$ linear constraints and three semidefinite constraints. To solve the optimization problem (25) with the required accuracy $\iota$, the interior point method requires $\mathcal{O}\left(\log\left(\iota^{-1}\right)\sqrt{2N_a + 2N_b + 1}\right)$ iterations, and the number of arithmetic operations is $\mathcal{O}\left(N_a^6 + N_b^6 + \left(N_a + N_b\right)^6 + \left(8K+7\right)N_a^2 + \left(8K+6\right)N_b^2 + \left(8K+8\right)N_aN_b + 4K + 4\right)$ in each iteration. Here, the operator $\mathcal{O}\left(\cdot\right)$ is the worst case computation bound [39]. Dropping the lower order terms, we obtain the computation complexity of the SITR algorithm as

$$\mathcal{O}\left(2\log\left(\epsilon\right)\log\left(\iota\right)\sqrt{2N_a + 2N_b + 1}\left(N_a^6 + N_b^6 + \left(N_a + N_b\right)^6\right)\right) \tag{26}$$

where $\epsilon$ is the required accuracy of the golden-section search in each iteration.

## IV. FA-SITR Maximization

Since the FA-SITR-maximization problem (21) shares the same feasible region with the SITR-maximization problem (18) when $\boldsymbol{W}_{a,b} = \boldsymbol{W}_{b,a} = \mathbf{0}$, the FA-SITR-maximization problem (21) is feasible if and only if the SITR-maximization problem (18) is feasible. Therefore, we can perform the feasibility check of the FA-SITR-maximization problem (21) via solving (22). The following analysis is based on the assumption that the optimization problem (21) is feasible.

Since the FA-SITR-maximization problem (21) is difficult to solve due to the non-convex objective function (21a) and the non-convex proportional fairness constraint in (21c). In order to deal with the non-convex proportional fairness constraint in (21c), we relax the equality of (21c) as

$$C_a\left(\boldsymbol{W}_{a,b}, \boldsymbol{W}_{b,a}, \boldsymbol{V}\right) \geq \phi_a \lambda \tag{27}$$

$$C_b\left(\boldsymbol{W}_{a,b}, \boldsymbol{W}_{b,a}, \boldsymbol{V}\right) \geq \phi_b \lambda \tag{28}$$

where $\lambda$ is a parameter that satisfies $\lambda > 0$. Note that the lower bounds of the information transmission rates $C_a\left(\boldsymbol{W}_{a,b}, \boldsymbol{W}_{b,a}, \boldsymbol{V}\right)$ and $C_b\left(\boldsymbol{W}_{a,b}, \boldsymbol{W}_{b,a}, \boldsymbol{V}\right)$ satisfy the proportional constraint. If the constraints in (27) and (28) are active[2], the proportional fairness constraint for the information transmission rates of the FD-UE $C_a\left(\boldsymbol{W}_{a,b}, \boldsymbol{W}_{b,a}, \boldsymbol{V}\right)$ and the FD-BST $C_b\left(\boldsymbol{W}_{a,b}, \boldsymbol{W}_{b,a}, \boldsymbol{V}\right)$ is satisfied.

In order to prove the activeness of the constraints in (27) and (28), we recast the FA-SITR-maximization problem (21) as follows

$$\max_{\boldsymbol{W}_{a,b}, \boldsymbol{W}_{b,a}, \boldsymbol{V}} \ \lambda \tag{29a}$$

$$\text{s.t. } (18\text{b}) - (18\text{i}) \tag{29b}$$

$$C_a\left(\boldsymbol{W}_{a,b}, \boldsymbol{W}_{b,a}, \boldsymbol{V}\right) \geq \phi_a \lambda \tag{29c}$$

$$C_b\left(\boldsymbol{W}_{a,b}, \boldsymbol{W}_{b,a}, \boldsymbol{V}\right) \geq \phi_b \lambda \tag{29d}$$

where $\lambda \in \left[0, \max\left\{\frac{1}{\phi_b}\log\left(1 + \frac{P_a^{\max}\sqrt{\mathrm{Tr}\left(\boldsymbol{H}_{a,b}\right)}}{\sigma_b^2}\right), \frac{1}{\phi_a}\log\left(1 + \frac{P_b^{\max}\sqrt{\mathrm{Tr}\left(\boldsymbol{H}_{b,a}\right)}}{\sigma_a^2}\right)\right\}\right]$.

***Proposition** 2:* Given $\lambda$, there exists a solution $\left(\widehat{\boldsymbol{W}}_{a,b}^{*}, \widehat{\boldsymbol{W}}_{b,a}^{*}, \widehat{\boldsymbol{V}}^{*}\right)$ such that the constraints in (29c) and (29d) are active.

[2]The term “active” indicates that a constraint is held with equality.

*Proof:* See Appendix C. □

From Proposition 2, we conclude that the proportional fairness constraint in (21c) can be satisfied. As a result, the objective value of the FA-SITR-maximization problem (21) is obtained via one-dimensional search for the optimal $\lambda$ and finding a feasible solution of (29) given $\lambda$. In addition, the system secrecy rate is obtained as $\lambda - \max\left\{C_a^{\text{LEAK}} + C_b^{\text{LEAK}}, C^{\text{LEAK}}\right\}$.

Dropping the rank-one constraints, we obtain the SDR version of the optimization problem (29) as

$$\max_{\boldsymbol{W}_{a,b},\boldsymbol{W}_{b,a},\boldsymbol{V}} \ \lambda \tag{30a}$$

$$\text{s.t.} \ \frac{\text{Tr}\left(\boldsymbol{H}_{b,e_k}\boldsymbol{W}_{b,a}\right)}{\exp\left(C_a^{\text{LEAK}}\right)-1} \leq \text{Tr}\left(\boldsymbol{H}_{a,e_k}\boldsymbol{W}_{a,b}\right) + \text{Tr}\left(\boldsymbol{H}_{e_k}\boldsymbol{V}\right) + \sigma_{e_k}^2, \forall k \tag{30b}$$

$$\frac{\text{Tr}\left(\boldsymbol{H}_{a,e_k}\boldsymbol{W}_{a,b}\right)}{\exp\left(C_b^{\text{LEAK}}\right)-1} \leq \text{Tr}\left(\boldsymbol{H}_{b,e_k}\boldsymbol{W}_{b,a}\right) + \text{Tr}\left(\boldsymbol{H}_{e_k}\boldsymbol{V}\right) + \sigma_{e_k}^2, \forall k \tag{30c}$$

$$\frac{\text{Tr}\left(\boldsymbol{H}_{a,e_k}\boldsymbol{W}_{a,b}\right) + \text{Tr}\left(\boldsymbol{H}_{b,e_k}\boldsymbol{W}_{b,a}\right)}{\exp\left(C^{\text{LEAK}}\right)-1} \leq \text{Tr}\left(\boldsymbol{H}_{e_k}\boldsymbol{V}\right) + \sigma_{e_k}^2, \forall k \tag{30d}$$

$$\frac{\text{Tr}\left(\boldsymbol{H}_{b,a}\boldsymbol{W}_{b,a}\right)}{\exp(\phi_a\lambda)-1} \geq \text{Tr}\left(\boldsymbol{H}_{a,a}\boldsymbol{W}_{a,b}\right) + \text{Tr}\left(\boldsymbol{H}_{a}\boldsymbol{V}\right) + \sigma_a^2 \tag{30e}$$

$$\frac{\text{Tr}\left(\boldsymbol{H}_{a,b}\boldsymbol{W}_{a,b}\right)}{\exp(\phi_b\lambda)-1} \geq \text{Tr}\left(\boldsymbol{H}_{b,b}\boldsymbol{W}_{b,a}\right) + \text{Tr}\left(\boldsymbol{H}_{b}\boldsymbol{V}\right) + \sigma_b^2 \tag{30f}$$

$$\text{Tr}\left(\boldsymbol{B}_a\boldsymbol{V}\right) + \text{Tr}\left(\boldsymbol{W}_{a,b}\right) \leq P_a^{\max} \tag{30g}$$

$$\text{Tr}\left(\boldsymbol{B}_b\boldsymbol{V}\right) + \text{Tr}\left(\boldsymbol{W}_{b,a}\right) \leq P_b^{\max} \tag{30h}$$

$$\text{Tr}\left(\boldsymbol{H}_{a,e_k}\boldsymbol{W}_{a,b}\right) + \text{Tr}\left(\boldsymbol{H}_{b,e_k}\boldsymbol{W}_{b,a}\right) + \text{Tr}\left(\boldsymbol{H}_{e_k}\boldsymbol{V}\right) + \sigma_{e_k}^2 \geq \frac{P_k^{\text{REQ}}}{\eta}, \forall k \tag{30i}$$

$$\boldsymbol{W}_{a,b} \succeq \boldsymbol{0}, \boldsymbol{W}_{b,a} \succeq \boldsymbol{0}, \boldsymbol{V} \succeq \boldsymbol{0}. \tag{30j}$$

***Proposition** 3:* Given $\lambda$, the solution $\left(\boldsymbol{W}_{a,b}^*, \boldsymbol{W}_{b,a}^*, \boldsymbol{V}^*\right)$ to (30) satisfies the rank-one constraints for $\boldsymbol{W}_{a,b}^*$ and $\boldsymbol{W}_{b,a}^*$, i.e., Rank $\left(\boldsymbol{W}_{a,b}^*\right) \leq 1$ and Rank $\left(\boldsymbol{W}_{b,a}^*\right) \leq 1$.

*Proof:* The proof follows similar arguments to those of Proposition 1, and we omit to avoid repetition. □

Based on Proposition 3, the optimal value of the optimization problem (21) can be obtained via: 1) solving the optimization problem (30); then 2) performing one-dimensional search for the optimal $\lambda^*$. We summarize the detailed procedure for solving the optimization problem (21) in Algorithm 2.

**Algorithm 2** FA-SITR Algorithm

1: FD-BST sets the iteration index $\tau = 0$ and the stop threshold $\epsilon$
2: FD-BST obtains $\lambda^{\min} = 0$ and $\lambda^{\max} = \max\left\{\frac{1}{\phi_b}\log\left(1 + P_a^{\max}\sqrt{\mathrm{Tr}\left(\boldsymbol{H}_{a,b}\right)}/\sigma_b^2\right), \frac{1}{\phi_a}\log\left(1 + P_b^{\max}\sqrt{\mathrm{Tr}\left(\boldsymbol{H}_{b,a}\right)}/\sigma_a^2\right)\right\}$
3: **repeat**
4: FD-BST obtains $\lambda = \frac{\lambda^{\min}+\lambda^{\max}}{2}$
5: FD-BST finds a feasible solution to the optimization problem (30)
6: **if** The optimization problem (30) is feasible given $\lambda$ **then**
7: $\lambda^{\max} \leftarrow \lambda$
8: **else**
9: $\lambda^{\min} \leftarrow \lambda$
10: **end if**
11: **until** $\left|\lambda^{\max} - \lambda^{\min}\right| \leq \epsilon$
12: **if** Rank $\left(\boldsymbol{W}^*_{a,b}\right) > 1$ and/or Rank $\left(\boldsymbol{W}^*_{b,a}\right) > 1$ **then**
13: FD-BST recovers the rank-one constraints via (60)-(62), and obtains $\widetilde{\boldsymbol{W}}^*_{a,b}$, $\widetilde{\boldsymbol{W}}^*_{b,a}$ and $\widetilde{\boldsymbol{V}}^*$.
14: **end if**
15: **if** The either (29c) or (29d) are inactive **then**
16: FD-BST recovers the activeness via (66)-(68), and obtains $\widehat{\boldsymbol{W}}^*_{a,b}$, $\widehat{\boldsymbol{W}}^*_{a,b}$, and $\widehat{\boldsymbol{V}}^*$
17: **end if**
18: FD-BST performs the eigenvalue decomposition for matrices $\widehat{\boldsymbol{W}}^*_{a,b}$, $\widehat{\boldsymbol{W}}^*_{b,a}$ and $\widehat{\boldsymbol{V}}^*$ in order to obtain the beamforming vectors $\boldsymbol{w}^*_{a,b}$ and $\boldsymbol{w}^*_{b,a}$ and the AN vectors $\boldsymbol{v}^*_a$ and $\boldsymbol{v}^*_b$.
19: FD-BST notifies the FD-UE the beamforming vector $\boldsymbol{w}^*_{b,a}$ and AN vector $\boldsymbol{v}^*_b$

*Complexity Analysis of the FA-SITR Algorithm:* Following similar arguments as in the complexity analysis of the SITR algorithm, we evaluate the computation complexity of the FA-SITR algorithm based on the complexity of the iteration loop from lines 3–11. In order to solve the optimization problem (30) via the interior point method, the number of iterations and the number of arithmetic operations are, respectively, counted as $\mathcal{O}\left(\log\left(\iota^{-1}\right)\sqrt{2N_a + 2N_b}\right)$ and $\mathcal{O}\left(N_a^6 + N_b^6 + \left(N_a + N_b\right)^6 + \left(8K + 7\right)\left(N_a^2 + N_b^2\right) + \left(8K + 8\right)N_aN_b\right)$. Dropping the lower order terms, we obtain the computation complexity of the FA-SITR algorithm as

$$\mathcal{O}\left(\log\left(\epsilon\right)\log\left(\iota\right)\sqrt{2N_a + 2N_b}\left(N_a^6 + N_b^6 + \left(N_a + N_b\right)^6\right)\right). \tag{31}$$

## V. Suboptimal Algorithms and Complexity Comparison

In this subsection, we propose two suboptimal solutions to each of the formulated problems: SITR-maximization and FA-SITR-maximization problems. In addition, we analyze the computation complexity of the proposed two suboptimal algorithms, and show that they enjoy lower complexity compared with the optimal algorithms in SITR Algorithm and FA-SITR Algorithm.

Our proposed two suboptimal algorithms are based on the setting that the covariance matrix $\boldsymbol{V}$ is in the null space of the matrix $[\boldsymbol{h}_a, \boldsymbol{h}_b]^{\mathrm{H}}$. Performing the singular value decomposition on the matrix $[\boldsymbol{h}_a, \boldsymbol{h}_b]^{\mathrm{H}}$, we have $[\boldsymbol{h}_a, \boldsymbol{h}_b]^{\mathrm{H}} = \boldsymbol{U}\boldsymbol{\Sigma}\boldsymbol{Y}^{\mathrm{H}} = \boldsymbol{U}\boldsymbol{\Sigma}[\widetilde{\boldsymbol{Y}}, \overline{\boldsymbol{Y}}]^{\mathrm{H}}$ where $\boldsymbol{U} \in \mathbb{C}^{2\times 2}$ and $\boldsymbol{Y} \in \mathbb{C}^{(N_a+N_b)\times(N_a+N_b)}$ are unitary matrices with $\boldsymbol{U}^{\mathrm{H}}\boldsymbol{U} = \boldsymbol{U}\boldsymbol{U}^{\mathrm{H}} = \boldsymbol{I}$ and $\boldsymbol{Y}^{\mathrm{H}}\boldsymbol{Y} = \boldsymbol{Y}\boldsymbol{Y}^{\mathrm{H}} = \boldsymbol{I}$. The matrix $\boldsymbol{\Sigma} \in \mathbb{C}^{2\times(N_a+N_b)}$ is a rectangular diagonal matrix. Here, the matrices $\widetilde{\boldsymbol{Y}} \in \mathbb{C}^{(N_a+N_b)\times 2}$ and $\overline{\boldsymbol{Y}} \in \mathbb{C}^{(N_a+N_b)\times(N_a+N_b-2)}$ are, respectively, the first two columns and the last $N_a + N_b - 2$ columns of the matrix $\boldsymbol{Y}$. Hence, the AN vectors can be obtained as

$$\boldsymbol{v}_n = \overline{\boldsymbol{Y}}\overline{\boldsymbol{v}}_n, n = 1, \ldots, N_a + N_b - 2 \tag{32}$$

where $\overline{\boldsymbol{v}}_n \in \mathbb{C}^{(N_a+N_b-2)\times 1}$ and $\|\overline{\boldsymbol{v}}_n\|_2 = 1$.

### A. LSI Nulling Algorithm

To remove the LSI from the transmission of the FD-BST and the FD-UE, the information beamforming vectors of the FD-BST and the FD-UE are selected as $\boldsymbol{h}_{a,a}^{\mathrm{H}}\boldsymbol{w}_{a,b} = 0$ and $\boldsymbol{h}_{b,b}^{\mathrm{H}}\boldsymbol{w}_{b,a} = 0$. Performing the singular value decomposition on $\boldsymbol{h}_{a,a}^{\mathrm{H}}$, we obtain $\boldsymbol{h}_{a,a}^{\mathrm{H}} = \boldsymbol{\Sigma}_{a,a}[\widetilde{\boldsymbol{X}}_{a,b}, \boldsymbol{X}_{a,b}]^{\mathrm{H}}$, where $\boldsymbol{\Sigma}_{a,a} \in \mathbb{C}^{1\times N_a}$, $\widetilde{\boldsymbol{X}}_{a,b} \in \mathbb{C}^{N_a\times 1}$ and $\boldsymbol{X}_{a,b} \in \mathbb{C}^{N_a\times(N_a-1)}$ with $[\widetilde{\boldsymbol{X}}_{a,b}, \boldsymbol{X}_{a,b}][\widetilde{\boldsymbol{X}}_{a,b}, \overline{\boldsymbol{X}}_{a,b}]^{\mathrm{H}} = [\widetilde{\boldsymbol{X}}_{a,b}, \boldsymbol{X}_{a,b}]^{\mathrm{H}}[\widetilde{\boldsymbol{X}}_{a,b}, \boldsymbol{X}_{a,b}] = \boldsymbol{I}$. Hence, the information beamforming vector of the FD-BST is obtained as

$$\boldsymbol{w}_{a,b} = \boldsymbol{X}_{a,b}\overline{\boldsymbol{w}}_{a,b} \tag{33}$$

where $\overline{\boldsymbol{w}}_{a,b} \in \mathbb{C}^{(N_a-1)\times 1}$ and $\|\overline{\boldsymbol{w}}_{a,b}\|_2 = 1$. Similarly, we obtain the information beamforming vector of the FD-UE as

$$\boldsymbol{w}_{b,a} = \boldsymbol{X}_{b,a}\overline{\boldsymbol{w}}_{b,a} \tag{34}$$

where $\overline{\boldsymbol{w}}_{b,a} \in \mathbb{C}^{(N_b-1)\times 1}$ and $\|\overline{\boldsymbol{w}}_{b,a}\|_2 = 1$. Here, the matrix $\boldsymbol{X}_{b,a}$ is obtained via the singular value decomposition on $\boldsymbol{h}_{b,b}^{\mathrm{H}} = \boldsymbol{\Sigma}_{b,b}[\widetilde{\boldsymbol{X}}_{b,a}, \boldsymbol{X}_{b,a}]^{\mathrm{H}}$ with $\boldsymbol{X}_{b,a} \in \mathbb{C}^{N_b\times(N_b-1)}$ and $\boldsymbol{X}_{b,a}^{\mathrm{H}}\boldsymbol{X}_{b,a} = \boldsymbol{I}$.

Applying the AN vectors (32) and the LSI null information beamforming vectors (33) and

(34), the SITR-maximization problem (18) reduces to

$$\max_{\overline{\boldsymbol{W}}_{a,b},\overline{\boldsymbol{W}}_{b,a},\overline{\boldsymbol{V}}} \log\left(1+\frac{\text{Tr}\left(\overline{\boldsymbol{H}}_{b,a}\overline{\boldsymbol{W}}_{b,a}\right)}{\sigma_a^2}\right)+\log\left(1+\frac{\text{Tr}\left(\overline{\boldsymbol{H}}_{a,b}\overline{\boldsymbol{W}}_{a,b}\right)}{\sigma_b^2}\right) \tag{35a}$$

$$\text{s.t.}\quad \frac{\text{Tr}\left(\overline{\boldsymbol{H}}_{b,e_k}\overline{\boldsymbol{W}}_{b,a}\right)}{\exp\left(C_a^{\text{LEAK}}\right)-1}\leq \text{Tr}\left(\overline{\boldsymbol{H}}_{a,e_k}\overline{\boldsymbol{W}}_{a,b}\right)+\text{Tr}\left(\overline{\boldsymbol{H}}_{e_k}\overline{\boldsymbol{V}}\right)+\sigma_{e_k}^2,\forall k \tag{35b}$$

$$\frac{\text{Tr}\left(\overline{\boldsymbol{H}}_{a,e_k}\overline{\boldsymbol{W}}_{a,b}\right)}{\exp\left(C_b^{\text{LEAK}}\right)-1}\leq \text{Tr}\left(\overline{\boldsymbol{H}}_{b,e_k}\overline{\boldsymbol{W}}_{b,a}\right)+\text{Tr}\left(\overline{\boldsymbol{H}}_{e_k}\overline{\boldsymbol{V}}\right)+\sigma_{e_k}^2,\forall k \tag{35c}$$

$$\frac{\text{Tr}\left(\overline{\boldsymbol{H}}_{a,e_k}\overline{\boldsymbol{W}}_{a,b}\right)+\text{Tr}\left(\overline{\boldsymbol{H}}_{b,e_k}\overline{\boldsymbol{W}}_{b,a}\right)}{\exp\left(C^{\text{LEAK}}\right)-1}\leq \text{Tr}\left(\overline{\boldsymbol{H}}_{e_k}\overline{\boldsymbol{V}}\right)+\sigma_{e_k}^2,\forall k \tag{35d}$$

$$\text{Tr}\left(\overline{\boldsymbol{B}}_a\overline{\boldsymbol{V}}\right)+\text{Tr}\left(\overline{\boldsymbol{W}}_{a,b}\right)\leq P_a^{\max} \tag{35e}$$

$$\text{Tr}\left(\overline{\boldsymbol{B}}_b\overline{\boldsymbol{V}}\right)+\text{Tr}\left(\overline{\boldsymbol{W}}_{b,a}\right)\leq P_b^{\max} \tag{35f}$$

$$\text{Tr}\left(\overline{\boldsymbol{H}}_{e_k}\overline{\boldsymbol{V}}\right)+\text{Tr}\left(\overline{\boldsymbol{H}}_{a,e_k}\overline{\boldsymbol{W}}_{a,b}\right)+\text{Tr}\left(\overline{\boldsymbol{H}}_{b,e_k}\overline{\boldsymbol{W}}_{b,a}\right)+\sigma_{e_k}^2\geq\frac{P_k^{\text{REQ}}}{\eta},\forall k \tag{35g}$$

$$\overline{\boldsymbol{W}}_{a,b}\succeq 0,\overline{\boldsymbol{W}}_{b,a}\succeq 0,\overline{\boldsymbol{V}}\succeq 0 \tag{35h}$$

$$\text{Rank}\left(\overline{\boldsymbol{W}}_{a,b}\right)\leq 1 \text{ and Rank}\left(\overline{\boldsymbol{W}}_{b,a}\right)\leq 1 \tag{35i}$$

where $\overline{\boldsymbol{H}}_{a,b}\triangleq\boldsymbol{X}_{a,b}^{\text{H}}\boldsymbol{H}_{a,b}\boldsymbol{X}_{a,b}$, $\overline{\boldsymbol{H}}_{b,a}\triangleq\boldsymbol{X}_{b,a}^{\text{H}}\boldsymbol{H}_{b,a}\boldsymbol{X}_{b,a}$, $\overline{\boldsymbol{H}}_{a,e_k}\triangleq\boldsymbol{X}_{a,b}^{\text{H}}\boldsymbol{H}_{a,e_k}\boldsymbol{X}_{a,b}$, $\overline{\boldsymbol{H}}_{b,e_k}\triangleq\boldsymbol{X}_{b,a}^{\text{H}}\boldsymbol{H}_{b,e_k}\boldsymbol{X}_{b,a}$, $\overline{\boldsymbol{H}}_{e_k}\triangleq\overline{\boldsymbol{Y}}^{\text{H}}\boldsymbol{H}_{e_k}\overline{\boldsymbol{Y}}$, $\overline{\boldsymbol{B}}_a=\overline{\boldsymbol{Y}}^{\text{H}}\boldsymbol{B}_a\overline{\boldsymbol{Y}}$ and $\overline{\boldsymbol{B}}_b=\overline{\boldsymbol{Y}}^{\text{H}}\boldsymbol{B}_b\overline{\boldsymbol{Y}}$. Here, the matrix $\overline{\boldsymbol{V}}=\sum_{n=1}^{N_a+N_b-2}\overline{\boldsymbol{v}}_n\overline{\boldsymbol{v}}_n^{\text{H}}$.

Using the SDR technique, the rank-one constraints in (35i) are recovered via similar procedures as in Proposition 1. On the other hand, we observe that the optimization problem in (35) contains three semidefinite matrix variables, $2+4K$ linear constraints and three positive semidefinite constraints. Therefore, the iterations required to solve (35) are $\mathcal{O}\left(\log\left(\iota^{-1}\right)\sqrt{2N_a+2N_b-4}\right)$, and the arithmetic operations in each iteration are $\mathcal{O}((N_a-1)^6+(N_b-1)^6+(N_a+N_b-2)^6+(4K+1)(N_a-1)^2+(4K+1)(N_b-1)^2+(4K+2)(N_a+N_b-2)^2)$. Dropping the lower order terms, we obtain the computation complexity of the LSI nulling (LSIN) algorithm as

$$\mathcal{O}\left(\log\left(\iota^{-1}\right)\sqrt{2N_a+2N_b-4}\left((N_a-1)^6+(N_b-1)^6+(N_a+N_b-2)^6\right)\right). \tag{36}$$

Similarly, applying the AN vectors (32) and the LSI nulling information beamforming vectors (33) and (34) to the FA-SITR maximization problem (21), we obtain the computation complexity of the fairness-aware LSIN (FA-LSIN) algorithm as

$$\mathcal{O}\left(\log\left(\epsilon\right)\log\left(\iota\right)\sqrt{2N_a+2N_b-4}\left((N_a-1)^6+(N_b-1)^6+(N_a+N_b-2)^6\right)\right). \tag{37}$$

### *B. LSI Nulling With Maximal Ratio Transmission Algorithm*

In order to reduce the computation complexity the LSI nulling algorithm, we assume that the information beamforming vectors are aligned to the channel coefficient vectors $\boldsymbol{h}_{a,b}$ and $\boldsymbol{h}_{b,a}$, i.e., maximal ratio transmission (MRT). In this case, the information beamforming vectors of the FD-BST and the FD-UE are, respectively, obtained as

$$\boldsymbol{w}_{a,b} = \sqrt{p_{a,b}} \frac{\boldsymbol{X}_{a,b}\boldsymbol{X}_{a,b}^{\mathrm{H}}\boldsymbol{h}_{a,b}}{\left\|\boldsymbol{X}_{a,b}^{\mathrm{H}}\boldsymbol{h}_{a,b}\right\|_2} \tag{38}$$

and

$$\boldsymbol{w}_{b,a} = \sqrt{p_{b,a}} \frac{\boldsymbol{X}_{b,a}\boldsymbol{X}_{b,a}^{\mathrm{H}}\boldsymbol{h}_{b,a}}{\left\|\boldsymbol{X}_{b,a}^{\mathrm{H}}\boldsymbol{h}_{b,a}\right\|_2} \tag{39}$$

where $p_{a,b}$ and $p_{b,a}$ are, respectively, the transmission power the FD-BST and the FD-UE.

Applying the AN vectors (32) and the information beamforming vectors (38) and (39), the SITR-maximization problem (18) reduces to

$$\max_{p_{a,b},p_{b,a},\overline{\boldsymbol{V}}} \log\left(1+\frac{p_{b,a}\left\|\boldsymbol{X}_{b,a}^{\mathrm{H}}\boldsymbol{h}_{b,a}\right\|_2^2}{\sigma_a^2}\right)+\log\left(1+\frac{p_{a,b}\left\|\boldsymbol{X}_{a,b}^{\mathrm{H}}\boldsymbol{h}_{a,b}\right\|_2^2}{\sigma_b^2}\right) \tag{40a}$$

$$\text{s.t. } p_{b,a}\frac{\left|\boldsymbol{h}_{b,e_k}^{\mathrm{H}}\boldsymbol{X}_{b,a}\boldsymbol{X}_{b,a}^{\mathrm{H}}\boldsymbol{h}_{b,a}\right|^2}{\left\|\boldsymbol{X}_{b,a}^{\mathrm{H}}\boldsymbol{h}_{b,a}\right\|_2^2\left(\exp\left(C_a^{\mathrm{LEAK}}\right)-1\right)} \le p_{a,b}\frac{\left|\boldsymbol{h}_{a,e_k}^{\mathrm{H}}\boldsymbol{X}_{a,b}\boldsymbol{X}_{a,b}^{\mathrm{H}}\boldsymbol{h}_{a,b}\right|^2}{\left\|\boldsymbol{X}_{a,b}^{\mathrm{H}}\boldsymbol{h}_{a,b}\right\|_2^2}+\mathrm{Tr}\left(\overline{\boldsymbol{H}}_{e_k}\overline{\boldsymbol{V}}\right)+\sigma_{e_k}^2, \forall k \tag{40b}$$

$$p_{a,b}\frac{\left|\boldsymbol{h}_{a,e_k}^{\mathrm{H}}\boldsymbol{X}_{a,b}\boldsymbol{X}_{a,b}^{\mathrm{H}}\boldsymbol{h}_{a,b}\right|^2}{\left\|\boldsymbol{X}_{a,b}^{\mathrm{H}}\boldsymbol{h}_{a,b}\right\|_2^2\left(\exp\left(C_b^{\mathrm{LEAK}}\right)-1\right)} \le p_{b,a}\frac{\left|\boldsymbol{h}_{b,e_k}^{\mathrm{H}}\boldsymbol{X}_{b,a}\boldsymbol{X}_{b,a}^{\mathrm{H}}\boldsymbol{h}_{b,a}\right|^2}{\left\|\boldsymbol{X}_{b,a}^{\mathrm{H}}\boldsymbol{h}_{b,a}\right\|_2^2}+\mathrm{Tr}\left(\overline{\boldsymbol{H}}_{e_k}\overline{\boldsymbol{V}}\right)+\sigma_{e_k}^2, \forall k \tag{40c}$$

$$p_{a,b}\frac{\left|\boldsymbol{h}_{a,e_k}^{\mathrm{H}}\boldsymbol{X}_{a,b}\boldsymbol{X}_{a,b}^{\mathrm{H}}\boldsymbol{h}_{a,b}\right|^2}{\left\|\boldsymbol{X}_{a,b}^{\mathrm{H}}\boldsymbol{h}_{a,b}\right\|_2^2\left(\exp\left(C^{\mathrm{LEAK}}\right)-1\right)} + p_{b,a}\frac{\left|\boldsymbol{h}_{b,e_k}^{\mathrm{H}}\boldsymbol{X}_{b,a}\boldsymbol{X}_{b,a}^{\mathrm{H}}\boldsymbol{h}_{b,a}\right|^2}{\left\|\boldsymbol{X}_{b,a}^{\mathrm{H}}\boldsymbol{h}_{b,a}\right\|_2^2\left(\exp\left(C^{\mathrm{LEAK}}\right)-1\right)}+\mathrm{Tr}\left(\overline{\boldsymbol{H}}_{e_k}\overline{\boldsymbol{V}}\right)+\sigma_{e_k}^2, \forall k \tag{40d}$$

$$\mathrm{Tr}\left(\overline{\boldsymbol{B}}_a\overline{\boldsymbol{V}}\right)+p_{a,b} \le P_a^{\max} \tag{40e}$$

$$\mathrm{Tr}\left(\overline{\boldsymbol{B}}_b\overline{\boldsymbol{V}}\right)+p_{b,a} \le P_b^{\max} \tag{40f}$$

$$\mathrm{Tr}\left(\overline{\boldsymbol{H}}_{e_k}\overline{\boldsymbol{V}}\right)+p_{a,b}\frac{\left|\boldsymbol{h}_{a,e_k}^{\mathrm{H}}\boldsymbol{X}_{a,b}\boldsymbol{X}_{a,b}^{\mathrm{H}}\boldsymbol{h}_{a,b}\right|^2}{\left\|\boldsymbol{X}_{a,b}^{\mathrm{H}}\boldsymbol{h}_{a,b}\right\|_2^2}+p_{b,a}\frac{\left|\boldsymbol{h}_{b,e_k}^{\mathrm{H}}\boldsymbol{X}_{b,a}\boldsymbol{X}_{b,a}^{\mathrm{H}}\boldsymbol{h}_{b,a}\right|^2}{\left\|\boldsymbol{X}_{b,a}^{\mathrm{H}}\boldsymbol{h}_{b,a}\right\|_2^2}+\sigma_{e_k}^2 \ge \frac{P_k^{\mathrm{REQ}}}{\eta}, \forall k \tag{40g}$$

$$p_{a,b} \ge 0, p_{b,a} \ge 0, \overline{\boldsymbol{V}} \succeq 0. \tag{40h}$$

We observe that the optimization problem (40) contains two scalar variables, one matrix variable, $4K+2$ linear constraints and one semidefinite constraint. Hence, $\mathcal{O}\left(\log\left(\iota^{-1}\right)\sqrt{N_a+N_b}\right)$ iterations and $\mathcal{O}\left(\left(N_a+N_b-2\right)^6+8K+4+\left(4K+2\right)\left(N_a+N_b-2\right)^2\right)$ arithmetic operations in each iteration are required to solve the optimization problem (40). Hence, the computation complexity of the LSI nulling with MRT (LSIN-MRT) algorithm for the SITR-maximization problem is given by

$$\mathcal{O}\left(\log\left(\iota^{-1}\right)\sqrt{N_a+N_b}\left(\left(N_a+N_b-2\right)^6\right)\right). \tag{41}$$

Similarly, the computation complexity of the fairness-aware LSIN-MRT (FA-LSIN-MRT) algorithm for the FA-SITR maximization problem is obtained as

$$\mathcal{O}\left(\log\left(\epsilon\right)\log\left(\iota\right)\sqrt{N_a+N_b}\left(\left(N_a+N_b-2\right)^6\right)\right). \tag{42}$$

Comparing the computation complexities of the proposed SITR algorithm in (26) and the two suboptimal algorithms (LSIN algorithm in (36) and LSIN-MRT algorithm in (41)), we observe that

$$\begin{aligned} &\underbrace{\mathcal{O}\left(2\log\left(\epsilon\right)\log\left(\iota\right)\sqrt{2N_a+2N_b+1}\left(N_a^6+N_b^6+\left(N_a+N_b\right)^6\right)\right)}_{\text{Complexity of SITR Algorithm}} \\ \geq &\underbrace{\mathcal{O}\left(\log\left(\iota^{-1}\right)\sqrt{2N_a+2N_b-4}\left(\left(N_a-1\right)^6+\left(N_b-1\right)^6+\left(N_a+N_b-2\right)^6\right)\right)}_{\text{Complexity of LSIN Algorithm}} \\ \geq &\underbrace{\mathcal{O}\left(\log\left(\iota^{-1}\right)\sqrt{N_a+N_b}\left(\left(N_a+N_b-2\right)^6\right)\right)}_{\text{Complexity of LSIN-MRT Algorithm}} \end{aligned} \tag{43}$$

where the two proposed suboptimal algorithms (LSIN algorithm and LSIN-MRT algorithm) have lower complexity than the SITR algorithm.

Comparing the computation complexities of the proposed FA-SITR algorithm in (31) and the two suboptimal algorithms (FA-LSIN algorithm in (37) and FA-LSIN-MRT algorithm in (42)), we conclude that the proposed two suboptimal algorithm (FA-LSIN algorithm and FA-LSIN-MRT algorithm) have lower complexity than the FA-SITR algorithm.

## VI. Numerical Results

In this section, we present numerical results to demonstrate the effectiveness of the proposed SITR and FA-SITR algorithms. Unless otherwise specified, the simulation parameters are set according to Table I.

First, we illustrate that the optimal values of the information leakage rate of the FD-BST and the FD-UE, and the total information leakage rate can be obtained via a two-dimensional search. This is due to the fact that the system secrecy rate is defined as the difference of the sum information transmission rate and the term $\max\{C_a^{\text{LEAK}}+C_b^{\text{LEAK}},C^{\text{LEAK}}\}$, i.e., $C_a+C_b-\max\{C_a^{\text{LEAK}}+C_b^{\text{LEAK}},C^{\text{LEAK}}\}$. As shown in Fig. 1 and Fig. 2, the first search dimension is the value of the total information leakage $C^{\text{LEAK}}$, and the second search dimension is the ratio $\frac{C_a^{\text{LEAK}}}{C^{\text{LEAK}}}$. Moreover, $\frac{C_a^{\text{LEAK}}}{C^{\text{LEAK}}}+\frac{C_b^{\text{LEAK}}}{C^{\text{LEAK}}}=1$. From Fig. 1, we observe that the system secrecy rate is jointly concave in $C^{\text{LEAK}}$ and $\frac{C_a^{\text{LEAK}}}{C^{\text{LEAK}}}$. Hence, we conclude that the optimal system secrecy rate can be obtained via the proposed SITR algorithm and a two-dimensional search method. A similar

TABLE I

SIMULATION PARAMETERS SETTING

| **Parameters** | **Values** |
|---|---|
| Number of ERs, $K$ | 4 |
| Number of Antenna at FD-BST, $N_a$ | 3 |
| Number of Antenna at FD-UE, $N_b$ | 3 |
| Gaussian noise power | $-30$ dBm |
| BST-to-UE channel, Rayleigh fading | $-35$ dB |
| BST-to-ER channel, Rayleigh fading | $-20$ dB |
| UE-to-ER channel, Rayleigh fading | $-20$ dB |
| Self-interference channel, Rayleigh fading | $-20$ dB |
| Information leakage rate | $C^{\text{LEAK}} = 0.6$ nats/sec/Hz |
| | $C_a^{\text{LEAK}} = 0.3$ nats/sec/Hz |
| | $C_b^{\text{LEAK}} = 0.3$ nats/sec/Hz |
| Maximum TX power of FD-BST, $P_a^{\max}$ | 20 dBm |
| Maximum TX power of FD-UE, $P_b^{\max}$ | 20 dBm |
| Minimum required input power, $P_k^{\text{req}}$ | 2 dBm |
| Energy harvester efficiency, $\eta$ | 0.5 |

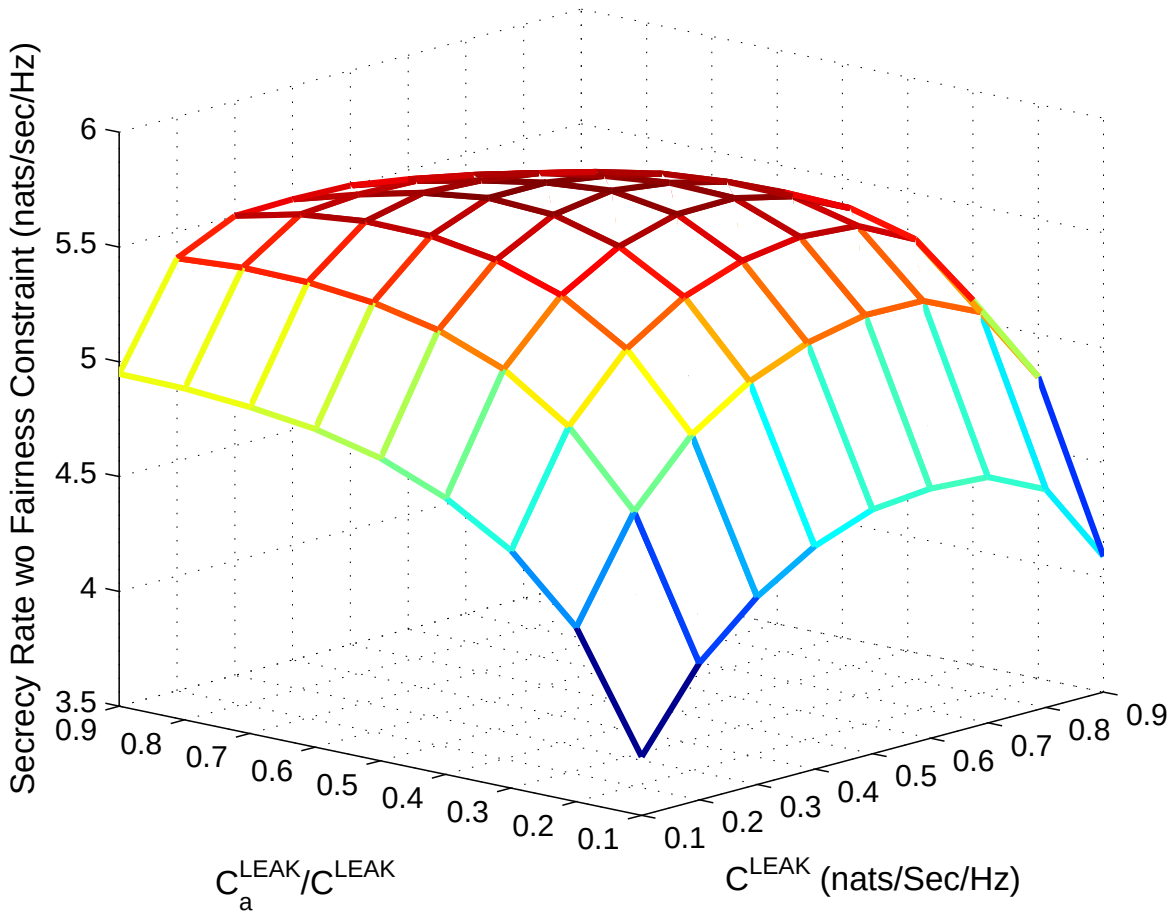

Fig. 1. An illustration of the two-dimensional search for the optimal system secrecy rate without the fairness constraint.

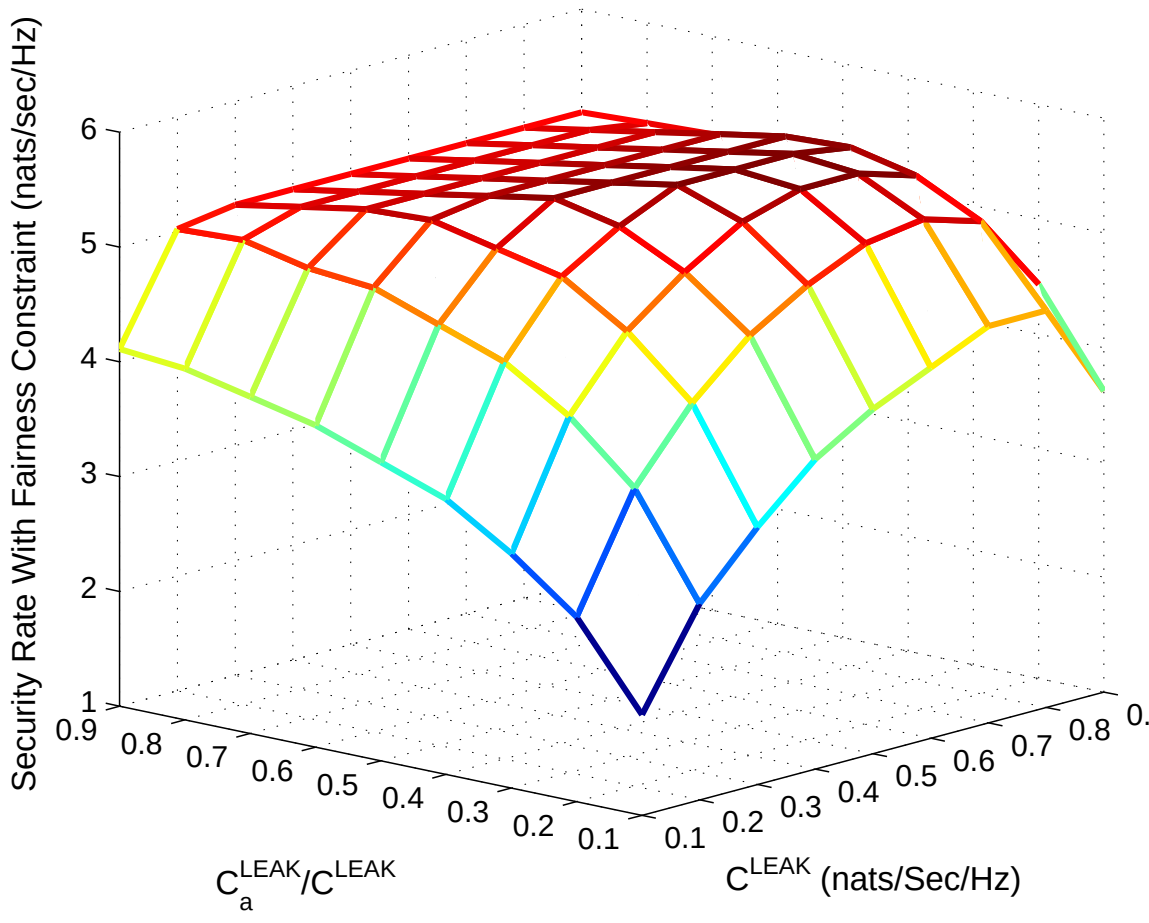

Fig. 2. An illustration of the two-dimensional search for the optimal system secrecy rate with the fairness constraint.

observation can be made from Fig. 2 that the fairness-aware optimal system secrecy rate can be obtained via the proposed FA-SITR algorithm and a two-dimensional search method.

Figure 3 shows the variation of the system secrecy rate with the required harvested energy of the ERs. We observe that the proposed SITR algorithm outperforms the two proposed suboptimal algorithms in the system secrecy rate. The SITR algorithm can improve the system secrecy rate over the LSIN algorithm by at most $1.78$ nats/sec/Hz, and over the LSIN-MRT algorithm by at most $1.87$ nats/sec/Hz. In addition, we observe that the system secrecy rate decreases with the required harvested energy. This is due to the fact that as more energy is consumed on the energy transmission at the FD-BST and the FD-UE, less energy is available for information transmission. When the fairness constraint in (21c) is considered, the proposed FA-SITR algorithm outperforms the two proposed suboptimal algorithms (LSIN algorithm and LSIN-MRT algorithm) as shown in Fig. 4. However, we also observe that the performance improvement of the proposed FA-SITR algorithm over FA-LSIN algorithm is not large. This observation can be explained as follows. The fairness constraint requires a sacrifice in the system information transmission rate of the proposed FA-SITR algorithm, FA-LSIN algorithm and FA-LSIN-MRT algorithm. For the FA-LSIN and FA-LSIN-MRT algorithms, the FD-BST and the FD-UE do not interfere with each other. Therefore, the fairness constraint can be achieved without much information transmission rate for the FA-LSIN and FA-LSIN-MRT algorithms.

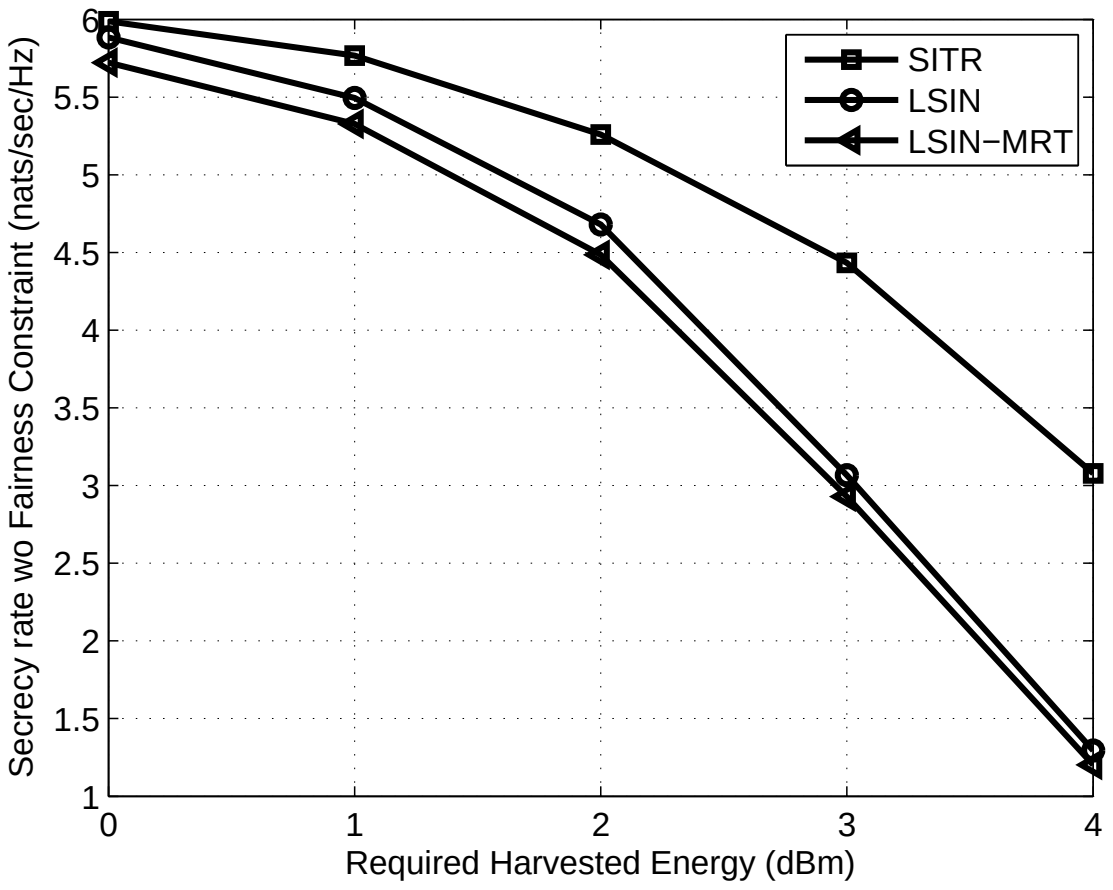

Fig. 3. The system secrecy rate versus the required harvested power without fairness constraint.

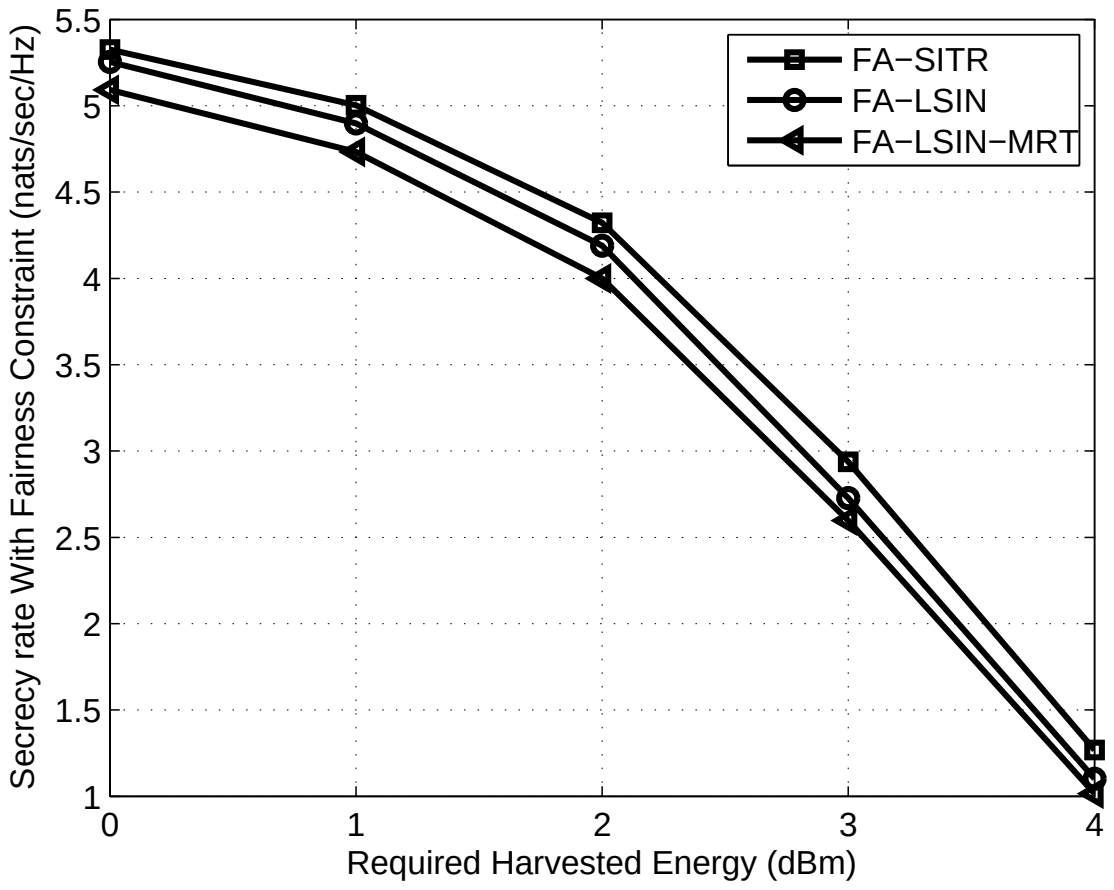

Fig. 4. The system secrecy rate versus the required harvested power with fairness constraint.

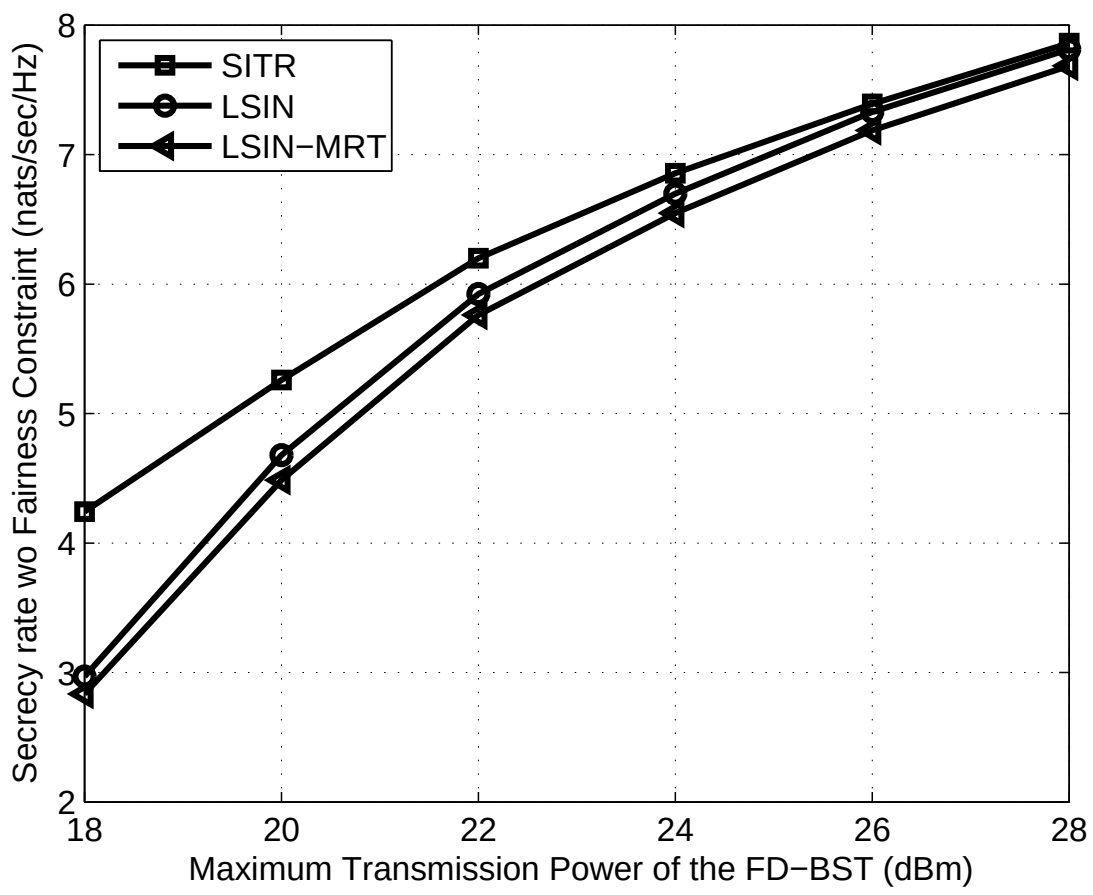

Fig. 5. The system secrecy rate versus maximum transmission power of the FD-BST without fairness constraint.

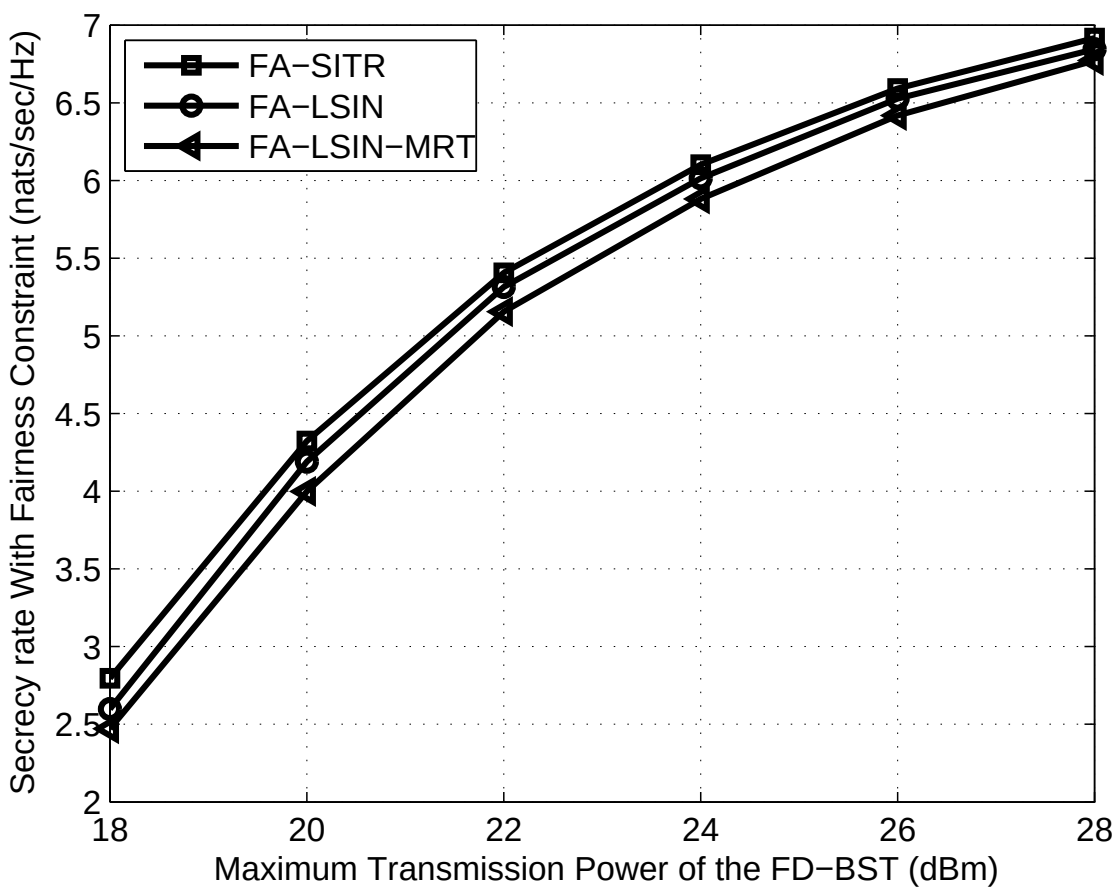

Fig. 6. The system secrecy rate versus maximum transmission power of the FD-BST with fairness constraint.

Figures 5 and 6 illustrate that the system secrecy rate varies with the maximum transmission power of the FD-BST. We observe that the system secrecy rate increases with the maximum transmission power of the FD-BST. Figure 5 shows that the performance gap between the proposed SITR algorithm and the proposed LSIN algorithm converges to zero as the maximum transmission power of the FD-BST increases. This is due to the fact that the large transmission power of the FD-BST can compensate for the loss of the degrees of the freedom of the multiple-input-multiple-output channels. From Fig. 5, we observe that the proposed LSIN-MRT algorithm achieves a near optimal secrecy rate when the transmission power of the FD-BST becomes large ($> 28$ dBm). Since the computation complexity of the proposed LSIN-MRT algorithm is significantly lower than that of the SITR algorithm, the LSIN-MRT algorithm is preferred when the transmission power of the FD-BST is large. When the fairness constraint is considered, the FS-SITR algorithm improves the system secrecy rate over the FA-LSIN algorithm and FA-LSIN-MRT algorithm by at most $7.6\%$ and $13.1\%$, respectively.

## VII. Conclusions

We investigated the SITR maximization and the FA-SITR maximization problems under constraints on the required harvested energy and the information leakage in FD-SWIPT systems. Information beamforming and AN vectors are designed jointly to solve the SITR and the FA-SITR maximization problems in FD-SWIPT systems. Combining a one-dimensional search method with the SDR technique, two optimal algorithms are developed for the SITR and the FA-SITR maximization problems. The optimality of the derived solutions is analytically proved. Numerical results also showed that the optimal system secrecy rate (i.e., secrecy capacity) can be obtained by introducing another two-dimensional search for the SITR and the FA-SITR algorithms. Finally, we proposed two suboptimal algorithms and showed numerically that one of them achieves a near-optimal secrecy rate to the SITR-maximization problem with an increasing maximum transmission power of the FD-BST.

## Appendix A
## Proof of Equivalence Between (23) and (25)

Assuming that $(\boldsymbol{W}_{a,b}, \boldsymbol{W}_{b,a}, \boldsymbol{V})$ is a solution to the optimization problem (23), we can construct a solution to the optimization problem (25) as follows

$$\overline{\gamma} = \frac{1}{\text{Tr}\left(\boldsymbol{H}_{a,b}\boldsymbol{W}_{a,b}\right) + \boldsymbol{h}_a^{\text{H}}\boldsymbol{V}\boldsymbol{h}_a + \sigma_a^2} \tag{44}$$

$$\overline{\boldsymbol{V}} = \frac{\boldsymbol{V}}{\text{Tr}\left(\boldsymbol{H}_{a,a}\boldsymbol{W}_{a,b}\right) + \boldsymbol{h}_a^{\text{H}}\boldsymbol{V}\boldsymbol{h}_a + \sigma_a^2} \tag{45}$$

$$\overline{\boldsymbol{W}}_{a,b} = \frac{\boldsymbol{W}_{a,b}}{\text{Tr}\left(\boldsymbol{H}_{a,a}\boldsymbol{W}_{a,b}\right) + \boldsymbol{h}_a^{\text{H}}\boldsymbol{V}\boldsymbol{h}_a + \sigma_a^2} \tag{46}$$

$$\overline{\boldsymbol{W}}_{b,a} = \frac{\boldsymbol{W}_{b,a}}{\text{Tr}\left(\boldsymbol{H}_{a,a}\boldsymbol{W}_{a,b}\right) + \boldsymbol{h}_a^{\text{H}}\boldsymbol{V}\boldsymbol{h}_a + \sigma_a^2}. \tag{47}$$

Substituting (44)-(47) into the optimization problem (25), we can verify that the constraints in (25b)-(25j) are satisfied. Moreover, the objective value of the optimization problem (23) is equal to that of the optimization problem (25). On the other hand, if $\left(\overline{\boldsymbol{W}}_{a,b}, \overline{\boldsymbol{W}}_{b,a}, \overline{\boldsymbol{V}}, \overline{\gamma}\right)$ is a solution to the optimization problem (25), $\left(\frac{\overline{\boldsymbol{W}}_{a,b}}{\gamma}, \frac{\overline{\boldsymbol{W}}_{b,a}}{\gamma}, \frac{\overline{\boldsymbol{V}}}{\gamma}\right)$ is a solution to the optimization problem (23).

## Appendix B
## Proof of Proposition 1

Introducing the dual variables $\mu, \nu \geq 0, \rho \geq 0, \tau \geq 0$ and $\{\theta_k \geq 0, \varpi_k \geq 0, \varrho_k \geq 0, \psi_k \geq 0\}_{k\in\mathcal{K}}$, the Lagrangian of the optimization problem (25) is given by

$$\begin{aligned} & L\left(\boldsymbol{W}_{a,b}, \boldsymbol{W}_{b,a}, \boldsymbol{V}, \mu, \nu, \rho, \tau, \{\theta_k, \varpi_k, \varrho_k, \psi_k\}_{k\in\mathcal{K}}\right) \\ & = -\mu + \text{Tr}\left(\boldsymbol{C}\boldsymbol{W}_{b,a}\right) + \text{Tr}\left(\boldsymbol{D}\boldsymbol{W}_{a,b}\right) + \text{Tr}\left(\boldsymbol{E}\boldsymbol{V}\right) + \varepsilon\gamma \end{aligned} \tag{48}$$

where

$$\boldsymbol{C} = \nu\boldsymbol{H}_{b,b} + \tau\boldsymbol{I} + \sum_{k=1}^{K}\left(\frac{\theta_k}{\exp\left(C_a^{\text{LEAK}}\right) - 1} + \frac{\varrho_k}{\exp\left(C^{\text{LEAK}}\right) - 1} - \varpi_k - \psi_k\right)\boldsymbol{H}_{b,e_k} - \boldsymbol{H}_{b,a} \tag{49}$$

$$\boldsymbol{E} = \mu\boldsymbol{H}_a + \nu\boldsymbol{H}_b + \rho\boldsymbol{B}_a + \tau\boldsymbol{B}_b - \sum_{k=1}^{K}\left(\theta_k + \varpi_k + \varrho_k + \psi_k\right)\boldsymbol{H}_{e_k} \tag{50}$$

$$\boldsymbol{D}=\mu\boldsymbol{H}_{a,a}+\rho\boldsymbol{I}+\sum_{k=1}^{K}\left(\frac{\varpi_k}{\exp\left(C_b^{\text{LEAK}}\right)-1}+\frac{\varrho_k}{\exp\left(C^{\text{LEAK}}\right)-1}-\theta_k-\psi_k\right)\boldsymbol{H}_{a,e_k}$$
$$-\frac{\nu}{\exp\left(\beta\right)-1}\boldsymbol{H}_{a,b} \tag{51}$$

$$\varepsilon=\mu\sigma_a^2+\nu\sigma_b^2-\rho P_a^{\max}-\tau P_b^{\max}-\sum_{k=1}^{K}\left(\theta_k+\varpi_k+\varrho_k\right)\sigma_{e_k}^2+\sum_{k=1}^{K}\psi_k\left(\frac{P_k^{\text{REQ}}}{\eta}-\sigma_{e_k}^2\right). \tag{52}$$

To guarantee the existence of the dual function, the Lagrangian (48) needs to be bounded below. Therefore, we have $\boldsymbol{C}^*\succeq\boldsymbol{0}$, $\boldsymbol{D}^*\succeq\boldsymbol{0}$, $\boldsymbol{E}^*\succeq\boldsymbol{0}$ and $\varepsilon^*\geq 0$ given the optimal value of dual variables $\mu^*$, $\nu^*$, $\rho^*$, $\tau^*$ and $\{\theta_k^*,\varpi_k^*,\varrho_k^*,\psi_k^*\}_{k\in\mathcal{K}}$. Hence, the dual problem of (25) is given by

$$\begin{aligned}\min_{\boldsymbol{W}_{a,b},\boldsymbol{W}_{b,a},\boldsymbol{V}}\ \max_{\substack{\mu,\nu,\rho,\tau\\ \{\theta_k,\varpi_k,\varrho_k,\psi_k\}_{k\in\mathcal{K}}}}\ &-\mu+\text{Tr}\left(\boldsymbol{C}\boldsymbol{W}_{b,a}\right)+\text{Tr}\left(\boldsymbol{D}\boldsymbol{W}_{a,b}\right)+\text{Tr}\left(\boldsymbol{E}\boldsymbol{V}\right)+\varepsilon\gamma\\ \text{s.t. }&\boldsymbol{C}\succeq\boldsymbol{0},\boldsymbol{D}\succeq\boldsymbol{0},\boldsymbol{E}\succeq\boldsymbol{0},\varepsilon\geq 0.\end{aligned} \tag{53}$$

Moreover, the Karush-Khun-Tucker (KKT) conditions related to the matrices $\boldsymbol{W}_{a,b}$, $\boldsymbol{W}_{b,a}$ and $\boldsymbol{V}$ are, respectively, given by

$$\boldsymbol{D}^*\boldsymbol{W}_{a,b}^*=\boldsymbol{0},\boldsymbol{C}^*\boldsymbol{W}_{b,a}^*=\boldsymbol{0}\text{ and }\boldsymbol{E}^*\boldsymbol{V}^*=\boldsymbol{0}. \tag{54}$$

Let $\boldsymbol{\Omega}^*=\mu^*\boldsymbol{H}_{a,a}+\rho^*\boldsymbol{I}+\sum_{k=1}^{K}\left(\frac{\varpi_k^*}{\exp\left(C_b^{\text{LEAK}}\right)-1}+\frac{\varrho_k^*}{\exp\left(C^{\text{LEAK}}\right)-1}-\theta_k^*-\psi_k^*\right)\boldsymbol{H}_{a,e_k}+\boldsymbol{H}_{a,b}$, we have

$$\boldsymbol{D}^*=\boldsymbol{\Omega}^*-\left(1+\frac{\nu^*}{\exp\left(\beta\right)-1}\right)\boldsymbol{H}_{a,b}. \tag{55}$$

To analyze the structure of the optimal matrix $\boldsymbol{W}_{a,b}^*$, we consider two cases for the matrix $\boldsymbol{\Omega}^*$.

- Case 1: Assuming that the matrix $\boldsymbol{\Omega}^*$ is full rank, that is, $\text{Rank}\left(\boldsymbol{\Omega}^*\right)=N_a$. Based on the KKT condition in (54) and (55) and the fact that $1+\frac{\nu^*}{\exp(\beta)-1}>0$, we obtain

$$\text{Rank}\left(\boldsymbol{W}_{a,b}^*\right)=\text{Rank}\left(\boldsymbol{\Omega}^*\boldsymbol{W}_{a,b}^*\right)=\text{Rank}\left(\left(1+\frac{\nu^*}{\exp\left(\beta\right)-1}\right)\boldsymbol{H}_{a,b}\boldsymbol{W}_{a,b}^*\right)\leq\text{Rank}\left(\boldsymbol{H}_{a,b}\right)=1. \tag{56}$$

  On the other hand, the optimal $\boldsymbol{W}_{a,b}^*$ cannot be the zero matrix since the minimum secure rate requirement $C_b^{\text{req}}$ is positive. We conclude that $\text{Rank}\left(\boldsymbol{W}_{a,b}^*\right)=1$.

- Case 2: Assuming that the matrix $\boldsymbol{\Omega}^*$ is rank deficient, that is, $\text{Rank}\left(\boldsymbol{\Omega}^*\right)=r_a<N_a$. Let $\boldsymbol{\Pi}=\left[\boldsymbol{\pi}_1,\ldots,\boldsymbol{\pi}_{N_a-r_a}\right]\in\mathbb{C}^{N_a\times(N_a-r_a)}$ denote the orthogonal basis matrix for the null space of the matrix $\mathbf{\Omega}^*$. Therefore, we have

$$\begin{aligned}\boldsymbol{\pi}_k^{\mathrm{H}}\boldsymbol{D}^*\boldsymbol{\pi}_k &= \boldsymbol{\pi}_k^{\mathrm{H}}\left[\mathbf{\Omega}^* - \left(1+\frac{\nu^*}{\exp(\beta)-1}\right)\boldsymbol{H}_{a,b}\right]\boldsymbol{\pi}_k \\ &= -\left(1+\frac{\nu^*}{\exp(\beta)-1}\right)\boldsymbol{\pi}_k^{\mathrm{H}}\boldsymbol{H}_{a,b}\boldsymbol{\pi}_k \geq 0\end{aligned} \tag{57}$$

where $\boldsymbol{\pi}_k$ is the $k$-th column of the matrix $\mathbf{\Pi}$. Since $\boldsymbol{H}_{a,b} \succeq \mathbf{0}$ and $1+\frac{\nu^*}{\exp(\beta)-1} > 0$, we conclude that $\boldsymbol{\pi}_k^{\mathrm{H}}\boldsymbol{H}_{a,b}\boldsymbol{\pi}_k = 0$, $k = 1,\ldots,N_a - r_a$. Hence, $\boldsymbol{\pi}_k^{\mathrm{H}}\boldsymbol{D}^*\boldsymbol{\pi}_k = 0$ and $\mathrm{Rank}(\boldsymbol{D}^*) \leq N_a - (N_a - r) = r$. From (55), we obtain $r_a - 1 \leq \mathrm{Rank}(\boldsymbol{D}^*) \leq r_a$. If $\mathrm{Rank}(\boldsymbol{D}^*) = r_a$, the matrices $\mathbf{\Omega}^*$ and $\boldsymbol{D}^*$ share the same orthogonal basis of the null space. As a result, the optimal $\boldsymbol{W}_{a,b}^* = \sum_{k=1}^{N_a-r_a}\zeta_k\boldsymbol{\pi}_k\boldsymbol{\pi}_k^{\mathrm{H}}$ with $\zeta_k \geq 0$. Moreover, $\mathrm{Tr}\left(\boldsymbol{W}_{a,b}^*\boldsymbol{H}_{a,b}\right) = 0$ due to the fact that $\boldsymbol{\pi}_k\boldsymbol{H}_{a,b}\boldsymbol{\pi}_k^{\mathrm{H}} = 0$. In this case, no information can be transmitted to the FD-UE. As a result, $\mathrm{Rank}(\boldsymbol{D}^*) = r_a - 1$, and the orthogonal basis of the null space of the matrix $\boldsymbol{D}^*$ is given by $[\mathbf{\Pi}, \boldsymbol{\pi}_0]$ with $\|\boldsymbol{\pi}_0\|_2 = 1$ and $\boldsymbol{\pi}_0^{\mathrm{H}}\boldsymbol{\pi}_k = 0$, $k = 1,\ldots,N_a - r_a$. In addition, $\boldsymbol{\pi}_0^{\mathrm{H}}\Xi^*\boldsymbol{\pi}_0 = \left(1+\frac{\nu^*}{\exp(\beta)-1}\right)\boldsymbol{\pi}_0^{\mathrm{H}}\boldsymbol{H}_{a,b}\boldsymbol{\pi}_0 > 0$ since $1+\frac{\nu^*}{\exp(\beta)-1} > 0$ and $\boldsymbol{\pi}_0$ does not lie in the null space of $\Xi^*$. We can rewrite the optimal matrix $\boldsymbol{W}_{a,b}^*$ as

$$\boldsymbol{W}_{a,b}^* = \zeta_0\boldsymbol{\pi}_0\boldsymbol{\pi}_0^{\mathrm{H}} + \sum_{k=1}^{N_a-r_a}\zeta_k\boldsymbol{\pi}_k\boldsymbol{\pi}_k^{\mathrm{H}}, \text{where } \zeta_0 \geq 0. \tag{58}$$

With (58), the FD-BST is able to transmit information to the FD-UE.

If the matrix $\nu^*\boldsymbol{H}_{b,b} + \tau^*\boldsymbol{I} + \sum_{k=1}^{K}\left(\frac{\theta_k^*}{\exp(C_a^{\mathrm{LEAK}})-1} + \frac{\varrho_k^*}{\exp(C^{\mathrm{LEAK}})-1} - \varpi_k^* - \psi_k^*\right)\boldsymbol{H}_{b,e_k}$ is full rank, the optimal matrix $\boldsymbol{W}_{b,a}^*$ is rank one. Otherwise, the optimal matrix $\boldsymbol{W}_{b,a}^*$ is given by

$$\boldsymbol{W}_{b,a}^* = \varsigma_0\boldsymbol{\xi}_0\boldsymbol{\xi}_0^{\mathrm{H}} + \sum_{k=1}^{N_b-r_b}\varsigma_k\boldsymbol{\xi}_k\boldsymbol{\xi}_k^{\mathrm{H}} \tag{59}$$

where $\varsigma_k \geq 0$, $k = 0,1,\ldots,N_b - r_b$; the scalar $r_b$ and the matrix $[\boldsymbol{\xi}_0,\boldsymbol{\xi}_1,\ldots,\boldsymbol{\xi}_{N_b-r_b}] \in \mathbb{C}^{N_b\times(N_b-r_b+1)}$ denote, respectively, the rank and the orthogonal basis of the null space of the matrix $\nu^*\boldsymbol{H}_{b,b} + \tau^*\boldsymbol{I} + \sum_{k=1}^{K}\left(\frac{\theta_k^*}{\exp(C_a^{\mathrm{LEAK}})-1} + \frac{\varrho_k^*}{\exp(C^{\mathrm{LEAK}})-1} - \varpi_k^* - \psi_k^*\right)\boldsymbol{H}_{b,e_k}$.

If $\mathrm{Rank}\left(\boldsymbol{W}_{a,b}^*\right) > 1$ and/or $\mathrm{Rank}\left(\boldsymbol{W}_{b,a}^*\right) > 1$, we can construct a new solution as

$$\widetilde{\boldsymbol{W}}_{a,b}^* = \boldsymbol{W}_{a,b}^* - \sum_{k=1}^{N_a-r_a}\zeta_k\boldsymbol{\pi}_k\boldsymbol{\pi}_k^{\mathrm{H}} \tag{60}$$

$$\widetilde{\boldsymbol{W}}_{b,a}^* = \boldsymbol{W}_{b,a}^* - \sum_{k=1}^{N_b-r_b}\varsigma_k\boldsymbol{\xi}_k\boldsymbol{\xi}_k^{\mathrm{H}} \tag{61}$$

$$\widetilde{\boldsymbol{V}}^* = \boldsymbol{V}^* + \begin{bmatrix} \sum_{k=1}^{N_a-r_a} \zeta_k \boldsymbol{\pi}_k \boldsymbol{\pi}_k^{\mathrm{H}} & \sum_{k=1}^{N_a-r_a} \sum_{j=1}^{N_b-r_b} \sqrt{\zeta_j \varsigma_k} \boldsymbol{\pi}_j \boldsymbol{\xi}_k^{\mathrm{H}} \\ \sum_{k=1}^{N_a-r_a} \sum_{j=1}^{N_b-r_b} \sqrt{\zeta_j \varsigma_k} \boldsymbol{\xi}_k \boldsymbol{\pi}_j^{\mathrm{H}} & \sum_{k=1}^{N_b-r_b} \varsigma_k \boldsymbol{\xi}_k \boldsymbol{\xi}_k^{\mathrm{H}} \end{bmatrix} \tag{62}$$

$$\widetilde{\gamma}^* = \gamma^* \tag{63}$$

where $\mathrm{Rank}\left(\widetilde{\boldsymbol{W}}_{a,b}^*\right) \leq 1$ and $\widetilde{\boldsymbol{W}}_{a,b}^* \leq 1$.

We can verify the optimality of the constructed solution $\left(\widetilde{\boldsymbol{W}}_{a,b}^*, \widetilde{\boldsymbol{W}}_{b,a}^*, \widetilde{\boldsymbol{V}}^*, \widetilde{\gamma}^*\right)$ via the following procedures. Substituting (59)-(62) into (25b)-(25h), we have

$$\mathrm{Tr}\left(\boldsymbol{H}_{b,a}\widetilde{\boldsymbol{W}}_{b,a}^*\right) = \mathrm{Tr}\left[\boldsymbol{H}_{b,a}\boldsymbol{W}_{b,a}^*\right] \tag{64a}$$

$$\mathrm{Tr}\left(\boldsymbol{H}_{a,a}\widetilde{\boldsymbol{W}}_{a,b}^*\right) + \boldsymbol{h}_a^{\mathrm{H}}\widetilde{\boldsymbol{V}}\boldsymbol{h}_a + \widetilde{\gamma}^*\sigma_a^2 = \mathrm{Tr}\left(\boldsymbol{H}_{a,a}\boldsymbol{W}_{a,b}^*\right) + \boldsymbol{h}_a^{\mathrm{H}}\boldsymbol{V}^*\boldsymbol{h}_a + \gamma^*\sigma_a^2 = 1 \tag{64b}$$

$$\max\left\{\frac{\mathrm{Tr}\left(\boldsymbol{H}_{a,e_k}\widetilde{\boldsymbol{W}}_{a,b}^*\right)}{1-\exp\left(-C_a^{\mathrm{LEAK}}\right)}, \frac{\mathrm{Tr}\left(\boldsymbol{H}_{b,e_k}\widetilde{\boldsymbol{W}}_{b,a}^*\right)}{1-\exp\left(-C_b^{\mathrm{LEAK}}\right)}, \frac{\mathrm{Tr}\left(\boldsymbol{H}_{a,e_k}\widetilde{\boldsymbol{W}}_{a,b}^*\right)+\mathrm{Tr}\left(\boldsymbol{H}_{b,e_k}\widetilde{\boldsymbol{W}}_{b,a}^*\right)}{1-\exp\left(-C^{\mathrm{LEAK}}\right)}\right\} \tag{64c}$$

$$\leq \mathrm{Tr}\left(\boldsymbol{H}_{a,e_k}\widetilde{\boldsymbol{W}}_{a,b}^*\right) + \mathrm{Tr}\left(\boldsymbol{H}_{b,e_k}\widetilde{\boldsymbol{W}}_{b,a}^*\right) + \mathrm{Tr}\left(\boldsymbol{H}_{e_k}\widetilde{\boldsymbol{V}}^*\right) + \sigma_{e_k}^2$$

$$\frac{\mathrm{Tr}\left(\boldsymbol{H}_{a,b}\widetilde{\boldsymbol{W}}_{a,b}\right)}{\exp(\beta)-1} \geq \mathrm{Tr}\left(\boldsymbol{H}_{b,b}\widetilde{\boldsymbol{W}}_{b,a}\right) + \mathrm{Tr}\left(\boldsymbol{H}_b\widetilde{\boldsymbol{V}}\right) + \widetilde{\gamma}\sigma_b^2 \tag{64d}$$

$$\mathrm{Tr}\left(\boldsymbol{B}_a\widetilde{\boldsymbol{V}}^*\right) + \mathrm{Tr}\left(\widetilde{\boldsymbol{W}}_{a,b}^*\right) \leq \widetilde{\gamma}^* P_a^{\max} \tag{64e}$$

$$\mathrm{Tr}\left(\boldsymbol{B}_b\widetilde{\boldsymbol{V}}^*\right) + \mathrm{Tr}\left(\widetilde{\boldsymbol{W}}_{b,a}^*\right) \leq \widetilde{\gamma}^* P_b^{\max} \tag{64f}$$

$$\widetilde{\gamma}^*\left(\frac{P_k^{\mathrm{REQ}}}{\eta} - \sigma_{e_k}^2\right) \leq \mathrm{Tr}\left(\boldsymbol{H}_{a,e_k}\widetilde{\boldsymbol{W}}_{a,b}^*\right) + \mathrm{Tr}\left(\boldsymbol{H}_{b,e_k}\widetilde{\boldsymbol{W}}_{b,a}^*\right) + \boldsymbol{h}_{e_k}^{\mathrm{H}}\widetilde{\boldsymbol{V}}^*\boldsymbol{h}_{e_k}. \tag{64g}$$

Note that (64a) achieves that the $\left(\widetilde{\boldsymbol{W}}_{a,b}^*, \widetilde{\boldsymbol{W}}_{b,a}^*, \widetilde{\boldsymbol{V}}^*, \widetilde{\gamma}^*\right)$ obtains the same objective value as $\left(\boldsymbol{W}_{a,b}^*, \boldsymbol{W}_{b,a}^*, \boldsymbol{V}^*, \gamma^*\right)$. In addition, Eq. (64b)-(64g) indicate that $\left(\widetilde{\boldsymbol{W}}_{a,b}^*, \widetilde{\boldsymbol{W}}_{b,a}^*, \widetilde{\boldsymbol{V}}^*, \widetilde{\gamma}^*\right)$ is also a feasible solution to the optimization problem (25). With $\mathrm{Rank}\left(\widetilde{\boldsymbol{W}}_{a,b}^*\right) \leq 1$ and $\mathrm{Rank}\left(\widetilde{\boldsymbol{W}}_{b,a}^*\right) \leq 1$, we conclude that $\left(\widetilde{\boldsymbol{W}}_{a,b}^*, \widetilde{\boldsymbol{W}}_{b,a}^*, \widetilde{\boldsymbol{V}}^*, \widetilde{\gamma}^*\right)$ is an optimal solution to the optimization problem (23).

## APPENDIX C
## PROOF OF PROPOSITION 2

If the constraints in (29c) and (29d) are inactive, we can find another

$$\lambda' = \min\left\{C_a\left(\boldsymbol{W}_{a,b}^*, \boldsymbol{W}_{b,a}^*, \boldsymbol{V}^*\right), C_b\left(\boldsymbol{W}_{a,b}^*, \boldsymbol{W}_{b,a}^*, \boldsymbol{V}^*\right)\right\} > \lambda \tag{65}$$

with $\mathcal{OBJ}_{\text{PF}}(\lambda') = \mathcal{OBJ}_{\text{PF}}(\lambda)$. As a result, the objective in (29) is increased.

Next, assume that either (29c) or (29d) is inactive. Let the constraint in (29c) be inactive and the constraint in (29d) be active with the optimal solution $\left(\boldsymbol{W}_{a,b}^*, \boldsymbol{W}_{b,a}^*, \boldsymbol{V}^*\right)$ to (29). We construct the solution $\left(\widehat{\boldsymbol{W}}_{a,b}^*, \widehat{\boldsymbol{W}}_{a,b}^*, \widehat{\boldsymbol{V}}^*\right)$ by the following procedures. There exists $\Delta \boldsymbol{W}_{b,a} \succeq \mathbf{0}$ such that

$$\widehat{\boldsymbol{W}}_{a,b}^* = \boldsymbol{W}_{a,b}^* \succeq \mathbf{0} \tag{66}$$

$$\widehat{\boldsymbol{W}}_{b,a}^* = \boldsymbol{W}_{b,a}^* - \Delta \boldsymbol{W}_{b,a} \succeq \mathbf{0} \tag{67}$$

$$\widehat{\boldsymbol{V}}^* = \boldsymbol{V}^* + \begin{bmatrix} \mathbf{0}_{N_a \times N_b} & \mathbf{0}_{N_a \times N_b} \\ \mathbf{0}_{N_b \times N_a} & \Delta \boldsymbol{W}_{b,a} \end{bmatrix} \succeq \mathbf{0} \tag{68}$$

such that

$$\frac{\text{Tr}\left(\boldsymbol{H}_{b,a}\widehat{\boldsymbol{W}}_{b,a}\right)}{\exp\left(\phi_a \lambda\right)} = \text{Tr}\left(\boldsymbol{H}_{a,a}\widehat{\boldsymbol{W}}_{a,b}\right) + \text{Tr}\left(\boldsymbol{H}_a \widehat{\boldsymbol{V}}\right) + \sigma_a^2 \tag{69}$$

$$\frac{\text{Tr}\left(\boldsymbol{H}_{a,b}\widehat{\boldsymbol{W}}_{a,b}\right)}{\exp\left(\phi_a \lambda\right)} = \text{Tr}\left(\boldsymbol{H}_{b,b}\widehat{\boldsymbol{W}}_{b,a}\right) + \text{Tr}\left(\boldsymbol{H}_b \widehat{\boldsymbol{V}}\right) + \sigma_b^2. \tag{70}$$

Substituting (66)-(68) into (18b)-(18g), we have

$$\max\left\{\frac{\text{Tr}\left(\boldsymbol{H}_{a,e_k}\widehat{\boldsymbol{W}}_{a,b}^*\right)}{1-\exp\left(-C_a^{\text{LEAK}}\right)}, \frac{\text{Tr}\left(\boldsymbol{H}_{b,e_k}\widehat{\boldsymbol{W}}_{b,a}^*\right)}{1-\exp\left(-C_b^{\text{LEAK}}\right)}, \frac{\text{Tr}\left(\boldsymbol{H}_{a,e_k}\widehat{\boldsymbol{W}}_{a,b}^*\right) + \text{Tr}\left(\boldsymbol{H}_{b,e_k}\widehat{\boldsymbol{W}}_{b,a}^*\right)}{1-\exp\left(-C^{\text{LEAK}}\right)}\right\} \tag{71a}$$

$$\leq \text{Tr}\left(\boldsymbol{H}_{a,e_k}\widehat{\boldsymbol{W}}_{a,b}^*\right) + \text{Tr}\left(\boldsymbol{H}_{b,e_k}\widehat{\boldsymbol{W}}_{b,a}^*\right) + \text{Tr}\left(\boldsymbol{H}_{e_k}\widehat{\boldsymbol{V}}^*\right) + \sigma_{e_k}^2$$

$$\text{Tr}\left(\boldsymbol{B}_a\widehat{\boldsymbol{V}}^*\right) + \text{Tr}\left(\widehat{\boldsymbol{W}}_{a,b}^*\right) \leq P_a^{\max} \tag{71b}$$

$$\text{Tr}\left(\boldsymbol{B}_b\widehat{\boldsymbol{V}}^*\right) + \text{Tr}\left(\widehat{\boldsymbol{W}}_{b,a}^*\right) \leq P_b^{\max} \tag{71c}$$

$$\text{Tr}\left(\boldsymbol{H}_{a,e_k}\widehat{\boldsymbol{W}}_{a,b}^*\right) + \text{Tr}\left(\boldsymbol{H}_{b,e_k}\widehat{\boldsymbol{W}}_{b,a}^*\right) + \text{Tr}\left(\boldsymbol{H}_{e_k}\widehat{\boldsymbol{V}}^*\right) + \sigma_{e_k}^2 \geq \frac{P_k^{\text{REQ}}}{\eta}. \tag{71d}$$

Based on (69), (70) and (71a)-(71d), we conclude that the solution $\left(\widehat{\boldsymbol{W}}_{a,b}^*, \widehat{\boldsymbol{W}}_{b,a}^*, \widehat{\boldsymbol{V}}^*\right)$ is also a solution to the optimization problem (29) that satisfy the proportional rate constraints.